\documentclass[aps,reprint,prl,superscriptaddress,floats,floatfix]{revtex4-2}

\usepackage{amsmath}
\usepackage{amsfonts}
\usepackage{amssymb}
\usepackage{graphicx}
\usepackage{color}
\usepackage{tikz}
\usepackage[colorlinks=true,citecolor=blue,linkcolor=blue,urlcolor=blue]{hyperref}
\usepackage{times}
\usepackage{amstext}
\usepackage{latexsym}
\usepackage[running,mathlines]{lineno}
\usepackage{bbm}
\usepackage{verbatim}
\usepackage{bm}

\usepackage{soul}
\usepackage{orcidlink}

\providecommand{\bra}[1]{\langle #1 \rvert}
\providecommand{\ket}[1]{\lvert #1 \rangle}

\providecommand{\ketbra}[2]{\lvert  #1\rangle \langle #2 \rvert}

\newcommand{\UFSCar}{Departamento de Física, Universidade Federal de São Carlos, Rodovia Washington Luís, km 235 - SP-310, 13565-905 São Carlos, SP, Brazil}

\begin{document}
\raggedbottom



\title{Dark States of Light and the Hidden Energy in Thermal Radiation Detection}

\author{Celso Jorge Villas-Boas~\orcidlink{0000-0001-5622-786X}}
\email{celsovb@df.ufscar.br}
\affiliation{\UFSCar}

\author{Ciro Micheletti~Diniz~\orcidlink{0000-0002-7602-0468}}
\email{mdciro@df.ufscar.br}
\affiliation{\UFSCar}


\begin{abstract}
We develop a quantum-optical framework demonstrating that thermal radiation can confine a significant portion of its energy in dark collective modes -- highly entangled photon states that, despite their photonic nature, remain decoupled from matter through standard electromagnetic interactions. In a system comprising $M$ thermal field modes, we show that only a fraction $1/M$ of the total energy is accessible to matter, while the remaining $(M-1)/M$ is stored in dark states, rendering it undetectable by conventional electromagnetic means. We also demonstrate that intensity measurements, commonly used to estimate field energy, can be misleading due to collective effects that suppress or enhance light-matter coupling. To explore further this phenomenon, we analyze a cavity QED model enclosing a single dissipative atom and show that symmetry breaking in the atom-field interaction enables access to the hidden energy stored in dark modes. While inconclusive, these findings suggest that dark states of light may underlie certain unexplained energy phenomena, pointing to a possible microscopic mechanism based on the collective structure of thermal radiation.

\end{abstract}

\maketitle

\section{Introduction}
Recently, the concept of collective dark states of light has been introduced, that is, states containing photons and energy yet not interacting with matter via electromagnetic coupling~\cite{VillasBoas2025}. This framework has offered new insights into the fundamental nature of light and interference phenomena, making it plausible, for instance, that photons may exist within the dark regions of double-slit experiments~\cite{VillasBoas2025}. When extended to superpositions involving $M$ modes, any pulse shape can be explained by bright and dark states of light. For instance, a mode-locked pulsed laser can be interpreted as a continuous photon flux partitioned into bright (pulse) and dark (inter-pulse) states~\cite{diniz2024}, operating in a regime dominated by dark states, i.e., for every single bright state, there are $M - 1$ corresponding dark states.

These dark states, involving two or more field modes, can host an arbitrarily large number of photons and, consequently, store significant amounts of energy. However, previous investigations focused either on single-photon states or coherent sources, where quantum effects are prominent. In contrast, most of the light in our daily environment appears as thermal radiation, prompting a natural question: Could this type of radiation also contain light in dark states? This hypothesis is particularly challenging, as thermal radiation is inherently incoherent, a feature that seemingly precludes the presence of dark states, which are highly non-classical and exhibit strong quantum entanglement~\cite{VillasBoas2025,Einstein1935, Horodecki2009}. 

In this manuscript, we demonstrate that thermal radiation contains projections onto collective dark modes -- or, equivalently, dark states -- and that the number of photons in such states increases with the number of field modes, as previously shown for coherent fields~\cite{diniz2024}. We formally prove this through two complementary approaches and reveal that a substantial portion of thermal energy resides in states that do not couple to matter. Specifically, for $M$ superposed modes at a given position, only a fraction $1/M$ of the initial thermal energy can be locally exchanged with matter, while the remaining portion, $(M-1)/M$, becomes inaccessible to it via electromagnetic interactions. We further show that standard intensity measurements can significantly misrepresent the actual energy content of multimode fields due to collective effects that enhance or suppress light-matter coupling. In fact, classical and quantum textbooks commonly consider field intensity as energy over units of area and time~\cite{jackson1999, loudon2000}, which is true for single-mode cases. However, for two or more modes, the intensity needs to be properly treated as the first-order correlation function~\cite{Glauber1963_1, Glauber1963, glauber2006}, which depends not only on the energy content but also on the coupling between the field and the detector, highlighting the importance of the careful treatment of the interaction process in order to obtain accurate understandings of the intensity measurements. Beyond the theoretical contributions, we show that all results can be experimentally tested using a simple and well-established setup in cavity quantum electrodynamics (QED) involving a single two-level atom trapped in a crossed-cavity system~\cite{brekenfeld2020}. This reinforces the significance of our findings and highlights their practical feasibility and experimental verifiability. Finally, we analyze the possibility that undetectable energy may exist in free space, pointing toward a potential explanation for certain forms of unknown energy.

\section{Model}
\begin{figure*}[t!]
\includegraphics[width=1.0\linewidth]{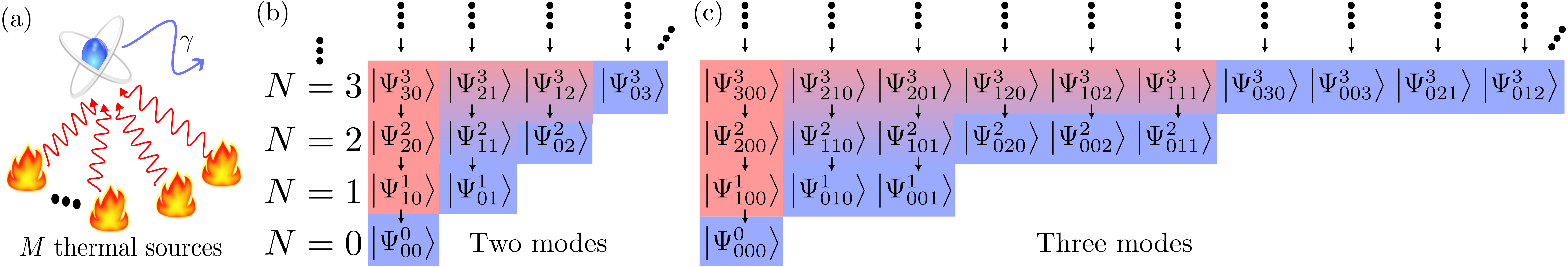}
	\caption{(a) Schematic representation of  $M$ thermal sources emitting in all directions. Their thermal photon beams may overlap at a given point in space, where they can interact with matter, here represented by a single dissipative atom (decay rate $\gamma$). Panels (b) and (c) illustrate the possible collective states for two- and three-field modes. For each total excitation number $N$, there exists a unique bright mode state (shown in red), denoted by $ \ket{\Psi^N_{N,0,\ldots,0}}$. In contrast, the number of dark mode states (shown in blue), denoted by $\ket{\Psi^N_{0,n_1,\ldots,n_{M-1}}}$, increases with both the number of modes $M$ and the excitation number $N$, where $ n_{\mu}$ is the number of photons in the $ {\mu} $-th dark mode. Finally, the states $ \ket{\Psi^N_{n_0,n_1,\ldots,n_{M-1}}} $, with $ n_0 \ne 0, N $ and $ \sum_{{\mu}=0}^{M-1} n_{\mu} = N $, represent intermediate configurations.}
	\label{Fig:Model}
\end{figure*}

To establish this quantum-optical framework, we begin by introducing the quantum model that fundamentally governs the exchange of energy between radiation and matter, considering 
$M$ thermal sources, as schematically illustrated in Fig.~\ref{Fig:Model}(a). For electronic excitations, this interaction is described by the Jaynes–Cummings model~\cite{Jaynes1963}, which can be derived from first principles, such as the minimal coupling theory, and by applying the rotating wave approximation (RWA)~\cite{scully1997, cohen1997}. For vibrational excitations, an analogous formalism applies, in which the interaction is mediated by the coupling between radiation and optical phonon modes in a given position in the material~\cite{Ciuti2005}. Thus, we consider the radiation–matter interaction described by the general Hamiltonian:
\begin{equation}
\hat{H}_\text{int} \propto \left( \hat{S}_+ \hat{E}^{(+)} + \hat{S}_- \hat{E}^{(-)} \right),
\label{JC}
\end{equation}
where $\hat{S}_+$ ($\hat{S}_- = \hat{S}_+^{\dagger}$) represents the excitation (de-excitation) of matter via the absorption (emission) of energy from (to) the radiation field. For electronic transitions, $\hat{S}_- = \hat{\sigma}_-$ is the lowering operator of a two-level system (related to the Pauli matrices), while for couplings involving optical phonons in a lattice, $\hat{S}_- = \hat{b}$ denotes the bosonic annihilation operator associated with the respective collective vibrational mode of the matter.

The operator $\hat{E}^{(+)}$ (with $\hat{E}^{(-)} = \hat{E}^{(+)\dagger}$) is the positive-frequency component of the electric field in a given position $\vec{r}$ and time $t$, which, according to Glauber’s formalism~\cite{Glauber1963, glauber2006, glauber2007quantum} and assuming linear polarization for simplicity, can be expressed as a sum over discrete field modes:
\begin{equation}
\hat{E}^{(+)} = \sum_{\vec{k}} g_{\vec{k}}(\vec{r},t)\hat{a}_{\vec{k}},
\end{equation}
where $\hat{a}_{\vec{k}}$ is the annihilation operator for the mode of frequency $\omega_{k}$ and wave vector $\vec{k}$, and the summation runs over $M$ independent modes. The functions $g_{\vec{k}}(\vec{r},t)$ depend on the system under study, but in general, they are proportional to $e^{-i(\vec{k} \cdot \vec{r} \pm \omega_k t)}$. Thus, they can induce distinct couplings and phases of the form $e^{i\theta_{\vec{k}}(\vec{r},t)}$, which may vary from point to point in spacetime. This is important because, at every position $\vec{r}$ and time $t$, there exists a set of collective modes (states) that are symmetric (bright) and antisymmetric (dark), as discussed below. For calculation purposes, these phases can be eliminated through a unitary transformation $U_{\bm{\theta}}$~\cite{diniz2024} while noting that the symmetric and antisymmetric modes ultimately retain a dependence on them. From this point onward, we work with the transformed positive-frequency electric field operator, $\hat{E}_{\bm{\theta}}^{(+)} = \sum_{k=1}^{M} \hat{a}_k$, which also assumes equal coupling strengths across all modes.

\section{Results and Discussions}

\subsection{Collective symmetric and antisymmetric modes} 
As shown in~\cite{VillasBoas2025}, when only a single mode is present, the vacuum state is the sole dark state of the electromagnetic field. However, for $M \geq 2$, the number of dark states increases with both the number of modes that effectively couple to matter and the total number of excitations in the system~\cite{VillasBoas2025,diniz2024}. To identify the states that do not couple to matter and are therefore unobservable, we change the basis from the bare mode representation to a collective mode basis. To this end, we start considering the discrete system of $M$ field modes described by the operators $\hat{a}_k$ ($k = 1,\ldots, M$), which satisfy the canonical commutation relations $[\hat{a}_j, \hat{a}_k^\dagger] = \delta_{jk}$. The system can be described either in the bare-mode basis or in terms of collective modes $\hat{A}_\mu$, defined via a unitary transformation~\cite{lemos2016,goldstein2002, Ponte2007, moussa2014, VillasBoas2025} (see also Supplementary Material (SM), which also includes the Refs.~\cite{diniz2024, Glauber1963_1, andrews1998theory, cameron1994combinatorics, Diniz_2025_reset, loudon2000, cohen1997, scully1997, eisberg1985, MadauDickinson2014, fermi1932, cohen1977, toledo_piza2023}):
\begin{equation}
\hat{A}_\mu = \sum_{j=1}^M U_{\mu j} \hat{a}_j, \quad \mu = 0, \ldots, M-1,
\label{unitary}
\end{equation}
where $U$ is a unitary matrix satisfying $\sum_{j=1}^M U_{\mu j} U_{\nu j}^* = \delta_{\mu \nu}$, which preserves the total number of excitations. A particularly relevant mode is the symmetric (or bright) collective mode, defined as
\begin{equation}
\hat{A}_0 = \frac{1}{\sqrt{M}} \sum_{j=1}^M \hat{a}_j.
\end{equation} 
The remaining $M-1$ collective modes, orthogonal to $\hat{A}_0$, are known as dark (or antisymmetric) modes. In terms of occupation numbers, a state with excitations in a given collective mode can be expressed in the bare basis via the inverse unitary transformation. Thus, excitations purely in the symmetric mode correspond to bright states, whereas those in orthogonal modes correspond to dark states. In the high-excitation limit, the collective-mode description becomes equivalent to the classical dynamics of coupled oscillators~\cite{Souza2015}, allowing for intuitive mechanical analogs~\footnote{Classical analogs as the symmetric and antisymmetric motions in coupled pendulum systems, familiar from undergraduate physics. In contrast, at low excitation levels, certain collective states exhibit genuinely nonclassical behavior that cannot be captured by classical theories~\cite{VillasBoas2025}.}. 

In this way, one can work either in the mode representation, as above, or in the state representation, as we perform in the Supplementary Material. For a total of $N$ excitations, the collective states are denoted as $\ket{\Psi^N_{n_0,n_1,\ldots,n_{M-1}}}$, where $n_{\mu}$ is the number of excitations in the $\mu$-th collective mode, with $\sum_{\mu=0}^{M-1} n_{\mu} = N$. When $n_0 = N$, all excitations reside in the symmetric (bright) mode; when $n_0 = 0$, they occupy only the antisymmetric (dark) modes~\cite{VillasBoas2025}.

In the collective basis, the matter, modeled here as a two-level or a bosonic system, interacts exclusively with the symmetric collective mode in such a way that the interaction Hamiltonian~(\ref{JC}) becomes (see SM):
\begin{equation}
\hat{H}_\text{int} \propto \sqrt{M} \left( \hat{S}_+ \hat{A}_0 + \hat{S}_- \hat{A}_0^\dagger \right),
\label{simetric_coupling}
\end{equation}
revealing an enhancement in the coupling strength between the symmetric mode only and the matter, due to collective effects. In contrast, the dark modes $\hat{A}_\mu$ ($\mu = 1, \ldots, M-1$) are orthogonal to $\hat{A}_0$ and therefore decoupled from matter. This follows from the fact that the coupling depends solely on the projection of the field onto the symmetric mode (bright states) and that all dark modes satisfy $\sum_{j=1}^M U_{\mu j} = 0$ for $\mu \geq 1$. As a result, any linear interaction with the matter necessarily vanishes for these modes. As previously discussed, introducing phase factors in the electric field operators can redefine the collective modes of the system. For a given set of relative phases, the matter may couple to a particular symmetric mode. However, altering these phases changes the structure of the collective modes, potentially converting an initially antisymmetric mode into the symmetric one -- thereby enabling its interaction with matter. This effect will be further examined in the context of symmetry breaking in the light-matter interaction between a single atom and two field modes within a cavity QED setting.

\subsection{Detectable Intensity \textit{vs} Energy}

The enhanced coupling of the symmetric mode and the decoupling of the antisymmetric modes introduce a subtle challenge for experimental interpretation. On the one hand, only the energy stored in the symmetric mode is effectively exchanged with matter; on the other hand, this exchange occurs at an enhanced rate, potentially producing a signal that may appear consistent with the total initial field energy, even though the actual energy transfer may be significantly smaller. This arises because typical measurements rely on field intensity, defined via the first-order correlation function~\cite{Glauber1963_1, Glauber1963, glauber2006}, which, for $M$ modes, yields $\langle \hat{E}^{(-)} \hat{E}^{(+)} \rangle = M \langle \hat{A}_0^\dagger \hat{A}_0 \rangle$. For a single mode $\hat{a}_1$, this quantity indeed is the average number of photons $\langle \hat{a}_1^{\dagger}\hat{a}_1 \rangle $ (energy), per area and time, no matter the light field state, and coincides with the usual definition of intensity as a function of the total energy of the field~\cite{jackson1999, loudon2000}. However, for two or more modes, the collective effects also reflect and highlight the field–matter coupling dependency in the intensity measurements, which can lead to situations where the intensity vanishes despite energy in the field (i.e., excitations in dark modes) or, conversely, where the measured intensity overestimates the actual energy content. For example, consider a single photon case: when it is in a single mode, the intensity returns $\langle \hat{E}^{(-)} \hat{E}^{(+)} \rangle = 1$; but, when the same single photon is delocalized over two modes ($\hat{a}_1$ and $\hat{a}_2$), prepared, for instance, in the state
\begin{equation}
    \ket{\Psi^1_{1,0}}=\frac{1}{\sqrt{2}} \left( \ket{1_{a_1},0_{a_2}} + \ket{0_{a_1},1_{a_2}} \right),
\end{equation}
the intensity is $\langle \hat{E}^{(-)} \hat{E}^{(+)} \rangle = 2$, that is twice that of a single-mode Fock state with one photon. This clearly illustrates that intensity measurements must not be interpreted as direct indicators of the field’s total energy. Rather, they reflect the rate of energy exchange -- i.e., the effective coupling strength. The situation becomes even more intriguing for mixed states. For example, consider the statistical mixture of single-photon states, over two modes,
\begin{equation}
    \hat{\rho}_{a_1,a_2} = \frac{1}{2} \left( \ket{1_{a_1},0_{a_2}} \bra{1_{a_1},0_{a_2}} + \ket{0_{a_1},1_{a_2}} \bra{0_{a_1},1_{a_2}} \right).
\end{equation}
The intensity calculation yields a value of $1$, corresponding to that state's correct energy. However, this result arises because only the fraction of energy stored in the symmetric mode (i.e., $1/2$) effectively contributes, but it does so with an enhancement factor of $2$ due to the collective effect in the interaction with the detector. Thus, direct intensity measurements in such cases yield values proportional to the full field energy, even though only a fraction of it is accessible through interaction with matter. In summary, direct measurements of the intensity of a radiation field in a thermal state will return a value associated with its energy. However, indirect measurements, for example, through light scattered by matter interacting with such thermal fields, can return values that are extremely discrepant in relation to the real energy values of the field in question. We explore this in more detail in the following sections, considering a practical example in a cavity QED setup. 

\subsection{Hidden energy in thermal states}

We are now ready to describe the energy exchange between matter and radiation and to identify the component that remains non-interacting, depending on the excitation configuration of the field. For an arbitrary field state with $N$ excitations, and assuming the matter is initially in the ground state $\ket{g}$, the interaction evolves the system as:
\begin{equation}
\hat{H}_\text{int} \ket{\Psi^N_{n_0,n_1,\ldots,n_{M-1}}} \ket{g} \propto \sqrt{M n_0} \ket{\Psi^{N-1}_{n_0 - 1, n_1,\ldots,n_{M-1}}} \ket{e},
\end{equation}
where $\ket{e}$ denotes the excited state of the matter. It is evident that when $n_0 = 0$ -- i.e., when there are no excitations in the symmetric mode -- no energy exchange takes place. Moreover, even when $n_0 > 0$, only a finite number of photons can be exchanged before the field transitions into a fully dark state with $N - n_0$ excitations, as exemplified in Fig.~\ref{Fig:Model}(b) and (c), for $M=2 \text{ and } 3$ modes, respectively.

Assuming that each bare mode $k$ is in a thermal state with average photon number $\bar{n}_k$, the density operator for each mode is given by
\begin{equation}
\hat{\rho}_k = \frac{1}{1+\bar{n}_k} \sum_{n=0}^\infty \left( \frac{\bar{n}_k}{1+\bar{n}_k} \right)^n |n\rangle_k \langle n|_k,
\end{equation}
where $|n\rangle_k$ are Fock states of bare mode $k$. Since the modes are independent, the total density operator is the tensor product
$\hat{\rho}_M = \bigotimes_{k=1}^M \hat{\rho}_k$.

To analyze how thermal energy is distributed among collective modes, let us first consider two uncorrelated thermal modes, $\hat{a}_1$ and $\hat{a}_2$. The symmetric ($\hat{A}_0$) and antisymmetric ($\hat{A}_1$) collective modes are, respectively,
\begin{equation}
\hat{A}_0 = \frac{\hat{a}_1 + \hat{a}_2}{\sqrt{2}}, \quad
\hat{A}_1 = \frac{\hat{a}_1 - \hat{a}_2}{\sqrt{2}},
\end{equation}
with number operators
$\hat{N}_0 = \hat{A}_0^\dagger \hat{A}_0$ and $\hat{N}_1 = \hat{A}_1^\dagger \hat{A}_1$.
Since the modes are uncorrelated and thermal, we have
\begin{equation}
\langle \hat{a}_j^\dagger \hat{a}_k \rangle = \delta_{jk} \bar{n}_j,
\end{equation}
leading to
\begin{equation}
\langle \hat{N}_0 \rangle = \langle \hat{N}_1 \rangle = \frac{\bar{n}_1 + \bar{n}_2}{2}.
\end{equation}
Thus, thermal energy is equally distributed between symmetric and antisymmetric modes. This result can be generalized to $M$ modes using the unitary transformation $U$ introduced in Eq.~(\ref{unitary}). With such transformation, one can easily calculate the average excitation number in the $\mu$-th collective mode, which reads
\begin{equation}
\langle \hat{N}_\mu \rangle = \sum_{j=1}^M |U_{\mu j}|^2 \bar{n}_j,
\end{equation}
since $\langle \hat{a}_j^\dagger \hat{a}_k \rangle = \delta_{jk} \bar{n}_j$ holds for any thermal, i.e., uncorrelated set of modes. For the symmetric mode,
\begin{equation}
U_{0 j} = \frac{1}{\sqrt{M}} \quad \Rightarrow \quad \langle \hat{N}_0 \rangle = \frac{1}{M} \sum_{j=1}^M \bar{n}_j.
\end{equation}
If all modes are identical, i.e., $\bar{n}_j = \bar{n}$, then thermal energy is uniformly distributed:
\begin{equation}
\langle \hat{N}_\mu \rangle = \bar{n}, \quad \forall \mu.
\end{equation}

We can now evaluate how the total field energy is distributed among the collective modes. The total energy, in the bare basis, is given by
$E_{\text{total}} = \sum_{j=1}^M \hbar \omega_j \bar{n}_j$,
and the energy associated with the symmetric mode is
\begin{equation}
E_S = \hbar \omega_S \langle \hat{N}_0 \rangle = \hbar \omega_S \frac{1}{M} \sum_{j=1}^M \bar{n}_j,
\end{equation}
being $\omega_S$ the frequency of the collective symmetric mode, which can be derived through the unitary transformation $U$ employed in Eq.~(\ref{unitary}).
The energy in the antisymmetric modes is then
\begin{equation}
E_A = E_{\text{total}} - E_S = \sum_{j=1}^M \hbar \omega_j \bar{n}_j - \hbar \omega_S \frac{1}{M} \sum_{j=1}^M \bar{n}_j.
\end{equation}
For identical modes, the collective modes become degenerate, i.e., $\omega_j = \omega_S = \omega$ and $\bar{n}_j = \bar{n}$, and then we finally find
\begin{equation}
E_A = \hbar \omega \left(1 - \frac{1}{M}\right) \sum_{j=1}^M \bar{n}_j = \hbar \omega \left( \frac{M-1}{M} \right) \sum_{j=1}^M \bar{n}.
\end{equation}
Hence, for large $M$, most of the thermal energy resides in the $M - 1$ antisymmetric modes~\footnote{This result assumes no coupling between modes, so the collective modes remain degenerate. When matter is included, we must guarantee that we are in the regime where the matter–mode coupling is much smaller than the mode frequencies, which justifies the rotating wave approximation (RWA). For stronger couplings (e.g., in the Rabi model), modes may become non-degenerate, and the properties of the collective modes change completely.}. Via a complementary approach, we derive in SM the same result by treating the photon distribution and the number of collective states as a combinatorial problem, considering the dependency of the number of collective states with the number of modes $M$ and total excitations $N$~\cite{diniz2024}. When interaction with matter is considered, only the symmetric mode couples to atoms, allowing them to absorb all of its energy. The antisymmetric modes remain unaffected, with their average excitations unchanged. Therefore, when many resonant modes exist in identical thermal states, only a fraction $1/M$ of the total thermal energy can be exchanged with matter. The remaining fraction, $(M-1)/M$, which can be substantial for large $M$, remains in dark modes (states), inaccessible via electromagnetic interaction.

\subsection{Experimental setup for proving hidden energy} 

In order to verify the previous results, consider only resonant modes and equal coupling strength $g$ between the bare modes and matter, for the sake of simplicity. Still, assuming that the energy exchanged does not flow back to the modes (e.g., radiation absorbed by a material surface and subsequently converted into heat), we can describe this process through a dissipative (decay rate $\gamma$) light-matter interaction in such a way that the system ($M$ modes + matter) evolves under the Born-Markov approximation, i.e., weak coupling and memoryless dynamics, which is ruled by the master equation~\cite{breuer2002,gardiner2004}:
\begin{equation}
\frac{d\hat{\rho}}{dt} = -\frac{i}{\hbar} [\hat{H}_{\text{int}}, \hat{\rho}] + \gamma \left( \hat{S}_- \hat{\rho} \hat{S}_+ - \frac{1}{2} \{ \hat{S}_+ \hat{S}_-, \hat{\rho} \} \right),
\label{master_equation}
\end{equation}
where $\hat{\rho}$ is the density operator of the entire system, comprising the matter and the $M$ radiation modes. The validity of this analysis relies on the assumption that all modes couple to the same matter element, here modeled as a single atom~\footnote{Should the modes interact with distinct material regions -- spaced by several wavelengths -- alternative symmetric states may arise, leading to spatially nonuniform energy transfer. Here, we are also neglecting the temperature effects of the matter since we are interested in the amount of energy of the radiation field that can be absorbed by the matter only.}. In the regime of weak light–matter coupling, i.e., $g \ll \gamma$, one can derive an effective master equation for the radiation field alone~\cite{Werlang2008, prado2009, prado2011}, which in the collective mode basis reads:
\begin{equation}
\frac{d\hat{\rho}_{R}}{dt} \simeq \kappa \left( \hat{A}_0 \hat{\rho}_{R} \hat{A}_0^{\dagger} - \frac{1}{2} \{ \hat{A}_0^{\dagger} \hat{A}_0, \hat{\rho}_{R} \} \right),
\label{master_efetiva}
\end{equation}
where $\hat{\rho}_R$ is the density matrix of the radiation field, and $\kappa = 4Mg^2/\gamma$~\cite{Werlang2008, prado2009, prado2011}. Again, this equation shows explicitly that only the symmetric collective mode will exchange energy with matter, while the antisymmetric modes remain decoupled. Therefore, if the initial energy is uniformly distributed among the bare modes, most of it becomes effectively ``dark'' to matter and unobservable. Even for thermal states, as discussed earlier, photon absorption leads the system toward a mixed state with the population concentrated in the dark modes.

\begin{figure}[t!]
    \centering
     \hspace*{-0.2cm} 
     \includegraphics[width=\columnwidth]{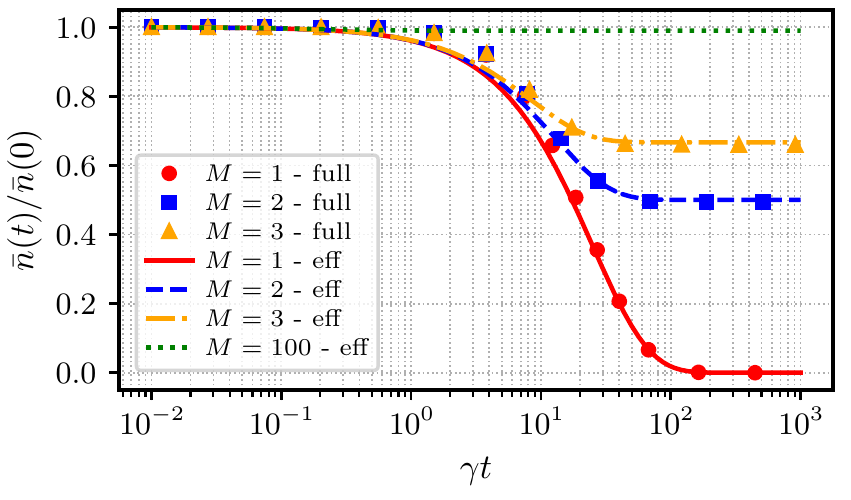}
    \caption{Time evolution of the average number of thermal photons distributed among $M$ modes, normalized by the initial value, $\bar{n}(t)/\bar{n}(0)$, for various values of $M$ (see legend). Lines correspond to the effective dynamics described by Eq.~\eqref{nt}, while symbols represent the full numerical solution of the master equation, Eq.~\eqref{master_equation}, considering $\gamma/g = 10$. As $M$ increases, the fraction of thermal radiation that becomes inaccessible to the matter subsystem grows markedly, approaching unity in the large-$M$ limit.}
    \label{fig:time_evolution_thermal_energy}
\end{figure}

The effective master equation above describes a standard dissipative bosonic mode whose dynamics are analytically solvable. For an initial thermal state, the average photon number in the symmetric mode evolves as $\langle \hat{A}_0^{\dagger} \hat{A}_0 \rangle(t) = \langle \hat{A}_0^{\dagger} \hat{A}_0 \rangle(0) e^{-\kappa t}$. As previously noted, for $M$ degenerate thermal modes, the initial mean photon number in the symmetric mode is $\bar{n}_S(0) = \bar{n}(0)/M$, while the antisymmetric modes contain $\bar{n}_A(0) = \bar{n}(0)(M-1)/M$, where $\bar{n}(0)$ is the total initial average photon number across all $M$ modes. Thus, the time evolution of the total average photon number is given by
\begin{equation}
\bar{n}(t) = \bar{n}_S(t) + \bar{n}_A(t) = \frac{1}{M} \bar{n}(0) e^{-\kappa t} + \frac{M-1}{M} \bar{n}(0).
\label{nt}
\end{equation}
This result demonstrates that, in the long-time limit, only the antisymmetric (dark) modes retain energy -- that is, energy that remains inaccessible to matter, even if it corresponds to a substantial number of photons. To illustrate this behavior, Fig.~\ref{fig:time_evolution_thermal_energy} shows the time evolution of the total average photon number, normalized by its initial value, as a function of $\gamma t$ for different values of $M$. The curves are obtained by numerically solving the full master equation~(\ref{master_equation}) using QuTiP~\cite{qutip5}, for $M = 1, 2$, and $3$, and are compared with the approximate analytical prediction given by Eq.~(\ref{nt}). As observed in the figure, for any $M$, the asymptotic photon ratios match the analytical results and align with the effective approach. For instance, $99\%$ of the initial energy remains confined to the dark modes when $M = 100$.

\begin{figure}[h!]
    \centering
     \hspace*{-0.2cm} 
     \includegraphics[width=\linewidth]{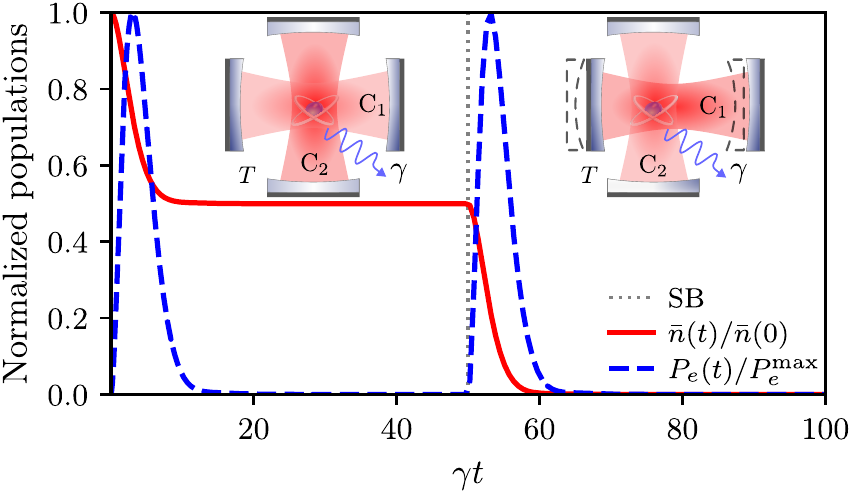}
    \caption{Normalized populations of the modes ($\bar{n}(t)/\bar{n}(0)$) and of the atom ($P_e(t)/P_e^{\mathrm{max}}$) as functions of $\gamma t$. Insets depict the experimental setup: two high-$Q$ cross-cavities ($\mathrm{C}_1$, $\mathrm{C}_2$) coupled to a dissipative atom (rate $\gamma$) via $g_{a_1}$ and $g_{a_2}$. Initially, $g_{a_1} = g_{a_2} = g$, and the system evolves to a steady state. Then, at $\gamma t = 50$, symmetry is broken (SB dotted line) by setting $g_{a_1} \rightarrow -g$, and the dynamics resumes. Both modes start in a thermal state with an average number of thermal photons $0.1$ (i.e., $\bar{n}(0) = 0.2$), and the atom is initially in the ground state. Initially, only half of the thermal photons are scattered, but after breaking symmetry, the atom scatters the remaining energy initially trapped in the dark mode. Results follow from numerical integration of Eq.~(\ref{master_equation}) with $\hat{S}_- = \hat{\sigma}_-$ and $|g| = \gamma/4$.}
    \label{fig:symmetry_breaking}
\end{figure}

The situation described above could be experimentally investigated using crossed-cavity setups~\cite{brekenfeld2020, farrera2020, hohmann2023, Solak2024, solak2024beamsplitter}, which allow for the observation of field states that do not interact with matter -- a form of hidden energy. Consider a dissipative atom coupled to two thermal modes of the electromagnetic field ($\hat{a}_1$ and $\hat{a}_2$), confined respectively in vertical and horizontal cavities -- see the left inset of Fig.~\ref{fig:symmetry_breaking}. Initially, the atom scatters light from the symmetric mode until it reaches the ground state, with the remaining energy trapped in the antisymmetric mode. However, as discussed earlier, by changing the phase of the electric field operators, the collective modes are also modified. Thus, by breaking the symmetry of the interaction -- for example, shifting the horizontal cavity so that $g_{a_1} = -g$ and $g_{a_2} = g$ -- the atom becomes coupled to the previous antisymmetric mode (dark state) and scatters the remaining energy, as shown in Fig.~\ref{fig:symmetry_breaking}, where we plot $\bar{n}(t)/\bar{n}(0)$ and the normalized atomic excited-state population $P_e(t)/P^{\mathrm{max}}_e$. While displacing cavity mirrors can be challenging, optical tweezers~\cite{seubert2025} offer a practical alternative to reposition the atom relative to the field modes to break the interaction symmetry. Alternatively, this symmetry breaking can occur dynamically through the introduction of different frequency detunings between the cavity modes and the atomic transition. These detunings continuously transform dark modes into bright ones and vice versa, enabling interaction with all modes and ultimately leading to the dissipation of the entire initial energy~\cite{Diniz_2025_reset} (see SM for more details).

\subsection{Non-detectable thermal  energy in free space}

In free space, the number of modes (at a given frequency and space-time point) can be estimated via the mode density per unit volume and unit frequency, given by $\mathcal{D}(\omega) = \omega^2/\pi^2 c^3$~\cite{loudon2000, cohen1997, scully1997}. Additionally, the matter can interact with modes whose frequencies are within a linewidth $\Gamma$, which is computed via Fermi's golden rule~\cite{fermi1932, cohen1977, toledo_piza2023}. Hence, the number of modes per volume unit that effectively are able to exchange energy with the matter is  $\mathcal{D}(\omega)\Gamma$. The larger this number is, the more collective effects play crucial roles during their interaction with the matter. Following what M. Planck did in 1900~\cite{eisberg1985}, we can compute the total thermal energy density present in all modes as $u_{\text{Total}} = \int_0^\infty \hbar\omega\, \mathcal{D}(\omega)\, \bar{n}(\omega)\, d\omega = \frac{\pi^2}{15} \cdot \frac{(k_B T)^4}{\hbar^3 c^3}$, considering an electromagnetic field in a thermal state, at temperature $T$, with an average photon number $\bar{n}(\omega) = 1 / (e^{\hbar \omega / k_B T} - 1)$, $k_B$ being the Boltzmann constant. Connecting to our model presented here, we can compare this energy density to the energy density resulting from a single mode per frequency (equivalently to assume the interaction only with a symmetric mode in a fixed frequency $\omega_S$) by considering the mode density in free space per unit volume and unit frequency as $\mathcal{D}_1(\omega=\omega_S) = 1$. Proceeding similarly as before, this density yields $u_1 = \int_0^\infty \hbar \omega_S\, \bar{n}(\omega_S)\, d\omega_S = \frac{\pi^2}{6} \cdot \frac{(k_B T)^2}{\hbar}$, and, consequently, an energy density ratio of $u_{\text{Total}}/u_1 = \frac{2}{5} \cdot \frac{(k_B T)^2}{\hbar^2 c^3}$. For an average stellar temperature $T\sim 3500$~K~\cite{MadauDickinson2014}, the ratio between the total energy density of thermal radiation and that associated with a symmetric mode is $u_{\text{Total}}/u_1 = 3117$. For comparison, at the Sun’s surface temperature, $T = 5770\,\text{K}$, the ratio increases to $u_{\text{Total}}/u_1 = 8471$. This means that, under stellar conditions (or the radiation from the radiation-matter decoupling era~\cite{planck2018}), the energy confined in non-interacting modes exceeds the observable field energy by more than three orders of magnitude. This excess energy, although invisible to matter, could be consistent with the observed gap between the radiation energy we detect and the estimated total energy content of the universe, commonly attributed to dark components~\cite{desy3, copeland2006, amendola2010,frieman2008,zwicky1933, bertone2005, planck2020}.

Critical aspects arise concerning what is measured in field intensity experiments because, depending on the specific characteristics of the detector, its response must be carefully analyzed. For instance, if radiation measurements are performed using sensors much smaller than the wavelength of the incoming radiation, the collective effects will be so strong that the sensor will interact only with the symmetric mode. In this way, only a fraction $1/M$ of the available energy couples to the symmetric mode, but with an enhanced coupling factor of $M$, which may lead to erroneous interpretations, although the well-controlled and direct measurement of the radiation intensity still represents the total energy. On the other hand, if the sensor dimensions are much larger than the radiation wavelength, the field phase will vary across the sensor surface, effectively giving rise to distinct symmetric modes at different points, thereby allowing for greater energy exchange with the radiation field.

There is also the question of indirect radiation measurements. For example, in indirect measurements of energy through scattered light by matter, e.g., from intergalactic areas that are dominated by hydrogen or helium atoms~\cite{shull2012}, atoms can absorb and scatter only a fraction $1/M$ of the total energy, despite doing so at an enhanced rate. Consequently, the energy they exchange and re-emit -- i.e., the observable signal -- corresponds to this reduced fraction. Thus, the energy scattered by atoms in the intergalactic medium would reflect only a small portion of the energy of the radiation field that actually surrounds them.

\subsection{Conclusion}

We have shown that thermal radiation can store a substantial fraction of its energy in dark collective modes, i.e., states that do not interact with matter through standard electromagnetic coupling. Even without coherence, thermal fields possess an intrinsic structure composed of symmetric (bright) and antisymmetric (dark) modes, or equivalently, bright and dark states, as formally derived here. In the limit of many modes, most of the energy is confined to dark states, making it inaccessible via linear interactions. However, this hidden component can be revealed through symmetry breaking in the light-matter coupling. These findings prompt a reconsideration of the classical picture of thermal radiation and may open new perspectives in quantum thermodynamics.

We have also shown that intensity measurements, described by first-order correlation functions~\cite{Glauber1963, Glauber1963_1, glauber2006}, may lead to misleading estimates of the actual energy present in field states. In the single-mode case ($\hat{a}_1$, frequency $\omega$), the intensity $\langle \hat{E}^{(-)} \hat{E}^{(+)}\rangle$ equals $\langle \hat{a}_1^\dagger \hat{a}_1 \rangle$, which corresponds to the total energy (in units of $\hbar\omega$). For multiple modes, however, the measured intensity accounts only for the energy in the symmetric (bright) component but is multiplied by the number of modes. This highlights a fundamental feature of intensity detection: it quantifies the strength of energy exchange with the detector, not the total field energy~\cite{Glauber1963_1, Glauber1963, glauber2006}. As a result, dark states, though energetically populated, yield zero intensity, while bright states may lead to an overestimation of the energy content. Thus, using intensity as a proxy for total energy must be approached with care, particularly in characterizing electromagnetic fields in free space or confined systems.

A particularly intriguing regime arises when thermal energy is uniformly distributed over $M$ modes. Although only a fraction $1/M$ of this energy is directly accessible to a detector, the collective coupling enhances the interaction strength by a $\sqrt{M}$ factor. Consequently, intensity measurements may yield energy values consistent with standard expectations while effectively probing only the symmetric portion of the total energy.

Numerical simulations of a single atom coupled to $M$ thermal modes support these findings: while the exchange rate is enhanced, only a fraction $1/M$ of the energy is absorbed and, subsequently, re-emitted. Accessing the remaining energy requires breaking the interaction symmetry, which can be experimentally achieved by displacing the atom~\cite{seubert2025} or creating different temporal dependencies between each mode and the atom~\cite{Diniz_2025_reset}. Similar techniques have been proposed in trapped-ion systems involving coherent or single-phonon excitations~\cite{parke2024phononic}, but not yet in the thermal regime.

The presence of substantial energy stored in multimode field states that remain undetectable to electromagnetic interactions, whether in single-photon, coherent, or thermal states~\cite{diniz2024}, raises the question of whether a significant fraction of energy might be distributed across space in dark states of light. As shown here, at any given spacetime point, only a small fraction of the available energy can excite atoms or vibrational modes of matter. This suggests the existence of energetically relevant components that are fundamentally inaccessible by standard electromagnetic means, potentially hinting at connections with forms of dark energy~\cite{desy3, copeland2006, amendola2010, frieman2008} or dark matter~\cite{zwicky1933, bertone2005, planck2020}. While this connection is intriguing, it is far from straightforward and would require a much deeper investigation, which lies beyond the scope of this work.

Finally, although our study is rooted in quantum electrodynamics (QED), the interaction model -- formulated in terms of creation and annihilation operators and excitation exchange in the second-quantization framework -- can be extended to other fundamental interactions, where analogous bright and dark states may also emerge~\cite{schonberg1952, schonberg1953a, schonberg1953b}.

\begin{acknowledgments}
\textit{Acknowledgments}. We would like to express our deepest gratitude for the support and inspiration from our friends and family (JC, NS, MVB, RVB, LVB, LFPD, EMD, REJS). We are also deeply thankful to Dario S. Thober, Paulo A. T. Muzy, and Salomon S. Mizrahi for their careful reading of this manuscript and for all the valuable criticisms and suggestions they provided. We also gratefully acknowledge the support from the São Paulo Research Foundation (FAPESP, Grants No. 2022/00209-6 and No. 2022/10218-2) and the Brazilian National Council for Scientific and Technological Development (CNPq, Grant No. 311612/2021-0).

\end{acknowledgments}

\bibliographystyle{apsrev4-2}
\bibliography{references.bib}

\begin{thebibliography}{55}%
\makeatletter
\providecommand \@ifxundefined [1]{%
 \@ifx{#1\undefined}
}%
\providecommand \@ifnum [1]{%
 \ifnum #1\expandafter \@firstoftwo
 \else \expandafter \@secondoftwo
 \fi
}%
\providecommand \@ifx [1]{%
 \ifx #1\expandafter \@firstoftwo
 \else \expandafter \@secondoftwo
 \fi
}%
\providecommand \natexlab [1]{#1}%
\providecommand \enquote  [1]{``#1''}%
\providecommand \bibnamefont  [1]{#1}%
\providecommand \bibfnamefont [1]{#1}%
\providecommand \citenamefont [1]{#1}%
\providecommand \href@noop [0]{\@secondoftwo}%
\providecommand \href [0]{\begingroup \@sanitize@url \@href}%
\providecommand \@href[1]{\@@startlink{#1}\@@href}%
\providecommand \@@href[1]{\endgroup#1\@@endlink}%
\providecommand \@sanitize@url [0]{\catcode `\\12\catcode `\$12\catcode `\&12\catcode `\#12\catcode `\^12\catcode `\_12\catcode `\%12\relax}%
\providecommand \@@startlink[1]{}%
\providecommand \@@endlink[0]{}%
\providecommand \url  [0]{\begingroup\@sanitize@url \@url }%
\providecommand \@url [1]{\endgroup\@href {#1}{\urlprefix }}%
\providecommand \urlprefix  [0]{URL }%
\providecommand \Eprint [0]{\href }%
\providecommand \doibase [0]{https://doi.org/}%
\providecommand \selectlanguage [0]{\@gobble}%
\providecommand \bibinfo  [0]{\@secondoftwo}%
\providecommand \bibfield  [0]{\@secondoftwo}%
\providecommand \translation [1]{[#1]}%
\providecommand \BibitemOpen [0]{}%
\providecommand \bibitemStop [0]{}%
\providecommand \bibitemNoStop [0]{.\EOS\space}%
\providecommand \EOS [0]{\spacefactor3000\relax}%
\providecommand \BibitemShut  [1]{\csname bibitem#1\endcsname}%
\let\auto@bib@innerbib\@empty
\bibitem [{\citenamefont {Villas-Boas}\ \emph {et~al.}(2025)\citenamefont {Villas-Boas}, \citenamefont {M\'aximo}, \citenamefont {Paulino}, \citenamefont {Bachelard},\ and\ \citenamefont {Rempe}}]{VillasBoas2025}%
  \BibitemOpen
  \bibfield  {author} {\bibinfo {author} {\bibfnamefont {C.~J.}\ \bibnamefont {Villas-Boas}}, \bibinfo {author} {\bibfnamefont {C.~E.}\ \bibnamefont {M\'aximo}}, \bibinfo {author} {\bibfnamefont {P.~J.}\ \bibnamefont {Paulino}}, \bibinfo {author} {\bibfnamefont {R.~P.}\ \bibnamefont {Bachelard}},\ and\ \bibinfo {author} {\bibfnamefont {G.}~\bibnamefont {Rempe}},\ }\href {https://doi.org/10.1103/PhysRevLett.134.133603} {\bibfield  {journal} {\bibinfo  {journal} {Phys. Rev. Lett.}\ }\textbf {\bibinfo {volume} {134}},\ \bibinfo {pages} {133603} (\bibinfo {year} {2025})}\BibitemShut {NoStop}%
\bibitem [{\citenamefont {Diniz}\ \emph {et~al.}(2024)\citenamefont {Diniz}, \citenamefont {Henrique}, \citenamefont {de~Souza}, \citenamefont {Misoguti}, \citenamefont {Ferreira},\ and\ \citenamefont {Villas-Boas}}]{diniz2024}%
  \BibitemOpen
  \bibfield  {author} {\bibinfo {author} {\bibfnamefont {C.~M.}\ \bibnamefont {Diniz}}, \bibinfo {author} {\bibfnamefont {F.~R.}\ \bibnamefont {Henrique}}, \bibinfo {author} {\bibfnamefont {B.~S.}\ \bibnamefont {de~Souza}}, \bibinfo {author} {\bibfnamefont {L.}~\bibnamefont {Misoguti}}, \bibinfo {author} {\bibfnamefont {P.~H.~D.}\ \bibnamefont {Ferreira}},\ and\ \bibinfo {author} {\bibfnamefont {C.~J.}\ \bibnamefont {Villas-Boas}},\ }\href@noop {} {\bibfield  {journal} {\bibinfo  {journal} {arXiv preprint arXiv:2412.19746}\ } (\bibinfo {year} {2024})}\BibitemShut {NoStop}%
\bibitem [{\citenamefont {Einstein}\ \emph {et~al.}(1935)\citenamefont {Einstein}, \citenamefont {Podolsky},\ and\ \citenamefont {Rosen}}]{Einstein1935}%
  \BibitemOpen
  \bibfield  {author} {\bibinfo {author} {\bibfnamefont {A.}~\bibnamefont {Einstein}}, \bibinfo {author} {\bibfnamefont {B.}~\bibnamefont {Podolsky}},\ and\ \bibinfo {author} {\bibfnamefont {N.}~\bibnamefont {Rosen}},\ }\href {https://doi.org/10.1103/PhysRev.47.777} {\bibfield  {journal} {\bibinfo  {journal} {Phys. Rev.}\ }\textbf {\bibinfo {volume} {47}},\ \bibinfo {pages} {777} (\bibinfo {year} {1935})}\BibitemShut {NoStop}%
\bibitem [{\citenamefont {Horodecki}\ \emph {et~al.}(2009)\citenamefont {Horodecki}, \citenamefont {Horodecki}, \citenamefont {Horodecki},\ and\ \citenamefont {Horodecki}}]{Horodecki2009}%
  \BibitemOpen
  \bibfield  {author} {\bibinfo {author} {\bibfnamefont {R.}~\bibnamefont {Horodecki}}, \bibinfo {author} {\bibfnamefont {P.}~\bibnamefont {Horodecki}}, \bibinfo {author} {\bibfnamefont {M.}~\bibnamefont {Horodecki}},\ and\ \bibinfo {author} {\bibfnamefont {K.}~\bibnamefont {Horodecki}},\ }\href {https://doi.org/10.1103/RevModPhys.81.865} {\bibfield  {journal} {\bibinfo  {journal} {Rev. Mod. Phys.}\ }\textbf {\bibinfo {volume} {81}},\ \bibinfo {pages} {865} (\bibinfo {year} {2009})}\BibitemShut {NoStop}%
\bibitem [{\citenamefont {Jackson}(1999)}]{jackson1999}%
  \BibitemOpen
  \bibfield  {author} {\bibinfo {author} {\bibfnamefont {J.~D.}\ \bibnamefont {Jackson}},\ }\href@noop {} {\emph {\bibinfo {title} {Classical Electrodynamics}}},\ \bibinfo {edition} {3rd}\ ed.\ (\bibinfo  {publisher} {John Wiley \& Sons},\ \bibinfo {address} {New York},\ \bibinfo {year} {1999})\BibitemShut {NoStop}%
\bibitem [{\citenamefont {Loudon}(2000)}]{loudon2000}%
  \BibitemOpen
  \bibfield  {author} {\bibinfo {author} {\bibfnamefont {R.}~\bibnamefont {Loudon}},\ }\href@noop {} {\emph {\bibinfo {title} {The Quantum Theory of Light}}},\ \bibinfo {edition} {3rd}\ ed.\ (\bibinfo  {publisher} {Oxford University Press},\ \bibinfo {address} {Oxford},\ \bibinfo {year} {2000})\BibitemShut {NoStop}%
\bibitem [{\citenamefont {Glauber}(1963{\natexlab{a}})}]{Glauber1963_1}%
  \BibitemOpen
  \bibfield  {author} {\bibinfo {author} {\bibfnamefont {R.~J.}\ \bibnamefont {Glauber}},\ }\href {https://doi.org/10.1103/PhysRev.130.2529} {\bibfield  {journal} {\bibinfo  {journal} {Phys. Rev.}\ }\textbf {\bibinfo {volume} {130}},\ \bibinfo {pages} {2529} (\bibinfo {year} {1963}{\natexlab{a}})}\BibitemShut {NoStop}%
\bibitem [{\citenamefont {Glauber}(1963{\natexlab{b}})}]{Glauber1963}%
  \BibitemOpen
  \bibfield  {author} {\bibinfo {author} {\bibfnamefont {R.~J.}\ \bibnamefont {Glauber}},\ }\href {https://doi.org/10.1103/PhysRev.131.2766} {\bibfield  {journal} {\bibinfo  {journal} {Phys. Rev.}\ }\textbf {\bibinfo {volume} {131}},\ \bibinfo {pages} {2766} (\bibinfo {year} {1963}{\natexlab{b}})}\BibitemShut {NoStop}%
\bibitem [{\citenamefont {Glauber}(2006)}]{glauber2006}%
  \BibitemOpen
  \bibfield  {author} {\bibinfo {author} {\bibfnamefont {R.~J.}\ \bibnamefont {Glauber}},\ }\href@noop {} {\bibfield  {journal} {\bibinfo  {journal} {Reviews of Modern Physics}\ }\textbf {\bibinfo {volume} {78}},\ \bibinfo {pages} {1267} (\bibinfo {year} {2006})}\BibitemShut {NoStop}%
\bibitem [{\citenamefont {Brekenfeld}\ \emph {et~al.}(2020)\citenamefont {Brekenfeld}, \citenamefont {Niemietz}, \citenamefont {Christesen},\ and\ \citenamefont {Rempe}}]{brekenfeld2020}%
  \BibitemOpen
  \bibfield  {author} {\bibinfo {author} {\bibfnamefont {M.}~\bibnamefont {Brekenfeld}}, \bibinfo {author} {\bibfnamefont {D.}~\bibnamefont {Niemietz}}, \bibinfo {author} {\bibfnamefont {J.~D.}\ \bibnamefont {Christesen}},\ and\ \bibinfo {author} {\bibfnamefont {G.}~\bibnamefont {Rempe}},\ }\href {https://doi.org/10.1038/s41567-020-0855-3} {\bibfield  {journal} {\bibinfo  {journal} {Nature Physics}\ ,\ \bibinfo {pages} {647}} (\bibinfo {year} {2020})}\BibitemShut {NoStop}%
\bibitem [{\citenamefont {Jaynes}\ and\ \citenamefont {Cummings}(1963)}]{Jaynes1963}%
  \BibitemOpen
  \bibfield  {author} {\bibinfo {author} {\bibfnamefont {E.}~\bibnamefont {Jaynes}}\ and\ \bibinfo {author} {\bibfnamefont {F.}~\bibnamefont {Cummings}},\ }\href {https://doi.org/10.1109/proc.1963.1664} {\bibfield  {journal} {\bibinfo  {journal} {Proceedings of the {IEEE}}\ }\textbf {\bibinfo {volume} {51}},\ \bibinfo {pages} {89} (\bibinfo {year} {1963})}\BibitemShut {NoStop}%
\bibitem [{\citenamefont {Scully}\ and\ \citenamefont {Zubairy}(1997)}]{scully1997}%
  \BibitemOpen
  \bibfield  {author} {\bibinfo {author} {\bibfnamefont {M.~O.}\ \bibnamefont {Scully}}\ and\ \bibinfo {author} {\bibfnamefont {M.~S.}\ \bibnamefont {Zubairy}},\ }\href@noop {} {\emph {\bibinfo {title} {Quantum Optics}}}\ (\bibinfo  {publisher} {Cambridge University Press},\ \bibinfo {year} {1997})\BibitemShut {NoStop}%
\bibitem [{\citenamefont {Cohen-Tannoudji}\ \emph {et~al.}(1997)\citenamefont {Cohen-Tannoudji}, \citenamefont {Dupont-Roc},\ and\ \citenamefont {Grynberg}}]{cohen1997}%
  \BibitemOpen
  \bibfield  {author} {\bibinfo {author} {\bibfnamefont {C.}~\bibnamefont {Cohen-Tannoudji}}, \bibinfo {author} {\bibfnamefont {J.}~\bibnamefont {Dupont-Roc}},\ and\ \bibinfo {author} {\bibfnamefont {G.}~\bibnamefont {Grynberg}},\ }\href@noop {} {\emph {\bibinfo {title} {Photons and Atoms: Introduction to Quantum Electrodynamics}}}\ (\bibinfo  {publisher} {Wiley},\ \bibinfo {year} {1997})\BibitemShut {NoStop}%
\bibitem [{\citenamefont {Ciuti}\ \emph {et~al.}(2005)\citenamefont {Ciuti}, \citenamefont {Bastard},\ and\ \citenamefont {Carusotto}}]{Ciuti2005}%
  \BibitemOpen
  \bibfield  {author} {\bibinfo {author} {\bibfnamefont {C.}~\bibnamefont {Ciuti}}, \bibinfo {author} {\bibfnamefont {G.}~\bibnamefont {Bastard}},\ and\ \bibinfo {author} {\bibfnamefont {I.}~\bibnamefont {Carusotto}},\ }\href {https://doi.org/10.1103/PhysRevB.72.115303} {\bibfield  {journal} {\bibinfo  {journal} {Physical Review B}\ }\textbf {\bibinfo {volume} {72}},\ \bibinfo {pages} {115303} (\bibinfo {year} {2005})}\BibitemShut {NoStop}%
\bibitem [{\citenamefont {Glauber}(2007)}]{glauber2007quantum}%
  \BibitemOpen
  \bibfield  {author} {\bibinfo {author} {\bibfnamefont {R.~J.}\ \bibnamefont {Glauber}},\ }\href@noop {} {\emph {\bibinfo {title} {Quantum theory of optical coherence: selected papers and lectures}}}\ (\bibinfo  {publisher} {John Wiley \& Sons},\ \bibinfo {year} {2007})\BibitemShut {NoStop}%
\bibitem [{\citenamefont {Lemos}(2016)}]{lemos2016}%
  \BibitemOpen
  \bibfield  {author} {\bibinfo {author} {\bibfnamefont {N.~A.}\ \bibnamefont {Lemos}},\ }\href@noop {} {\emph {\bibinfo {title} {Mecânica Analítica}}},\ \bibinfo {edition} {2nd}\ ed.\ (\bibinfo  {publisher} {Livraria da Física},\ \bibinfo {address} {São Paulo, Brasil},\ \bibinfo {year} {2016})\ p.\ \bibinfo {pages} {394},\ \bibinfo {note} {disponível em capa dura e mole}\BibitemShut {NoStop}%
\bibitem [{\citenamefont {Goldstein}\ \emph {et~al.}(2002)\citenamefont {Goldstein}, \citenamefont {Poole},\ and\ \citenamefont {Safko}}]{goldstein2002}%
  \BibitemOpen
  \bibfield  {author} {\bibinfo {author} {\bibfnamefont {H.}~\bibnamefont {Goldstein}}, \bibinfo {author} {\bibfnamefont {C.}~\bibnamefont {Poole}},\ and\ \bibinfo {author} {\bibfnamefont {J.}~\bibnamefont {Safko}},\ }\href@noop {} {\emph {\bibinfo {title} {Classical Mechanics}}},\ \bibinfo {edition} {3rd}\ ed.\ (\bibinfo  {publisher} {Addison-Wesley},\ \bibinfo {address} {San Francisco},\ \bibinfo {year} {2002})\ p.\ \bibinfo {pages} {672}\BibitemShut {NoStop}%
\bibitem [{\citenamefont {de~Ponte}\ \emph {et~al.}(2007)\citenamefont {de~Ponte}, \citenamefont {Mizrahi},\ and\ \citenamefont {Moussa}}]{Ponte2007}%
  \BibitemOpen
  \bibfield  {author} {\bibinfo {author} {\bibfnamefont {M.~A.}\ \bibnamefont {de~Ponte}}, \bibinfo {author} {\bibfnamefont {S.~S.}\ \bibnamefont {Mizrahi}},\ and\ \bibinfo {author} {\bibfnamefont {M.~H.~Y.}\ \bibnamefont {Moussa}},\ }\href {https://doi.org/10.1103/PhysRevA.76.032101} {\bibfield  {journal} {\bibinfo  {journal} {Phys. Rev. A}\ }\textbf {\bibinfo {volume} {76}},\ \bibinfo {pages} {032101} (\bibinfo {year} {2007})}\BibitemShut {NoStop}%
\bibitem [{\citenamefont {Máximo}\ \emph {et~al.}(2014)\citenamefont {Máximo}, \citenamefont {Batalhão}, \citenamefont {Bachelard}, \citenamefont {de~Moraes~Neto}, \citenamefont {de~Ponte},\ and\ \citenamefont {Moussa}}]{moussa2014}%
  \BibitemOpen
  \bibfield  {author} {\bibinfo {author} {\bibfnamefont {C.~E.}\ \bibnamefont {Máximo}}, \bibinfo {author} {\bibfnamefont {T.~B.}\ \bibnamefont {Batalhão}}, \bibinfo {author} {\bibfnamefont {R.}~\bibnamefont {Bachelard}}, \bibinfo {author} {\bibfnamefont {G.~D.}\ \bibnamefont {de~Moraes~Neto}}, \bibinfo {author} {\bibfnamefont {M.~A.}\ \bibnamefont {de~Ponte}},\ and\ \bibinfo {author} {\bibfnamefont {M.~H.~Y.}\ \bibnamefont {Moussa}},\ }\href@noop {} {\bibfield  {journal} {\bibinfo  {journal} {J. Opt. Soc. Am. B}\ }\textbf {\bibinfo {volume} {31}},\ \bibinfo {pages} {2480} (\bibinfo {year} {2014})}\BibitemShut {NoStop}%
\bibitem [{\citenamefont {Andrews}(1998)}]{andrews1998theory}%
  \BibitemOpen
  \bibfield  {author} {\bibinfo {author} {\bibfnamefont {G.~E.}\ \bibnamefont {Andrews}},\ }\href@noop {} {\emph {\bibinfo {title} {The theory of partitions}}},\ \bibinfo {number} {2}\ (\bibinfo  {publisher} {Cambridge university press},\ \bibinfo {year} {1998})\BibitemShut {NoStop}%
\bibitem [{\citenamefont {Cameron}(1994)}]{cameron1994combinatorics}%
  \BibitemOpen
  \bibfield  {author} {\bibinfo {author} {\bibfnamefont {P.~J.}\ \bibnamefont {Cameron}},\ }\href@noop {} {\emph {\bibinfo {title} {Combinatorics: topics, techniques, algorithms}}}\ (\bibinfo  {publisher} {Cambridge University Press},\ \bibinfo {year} {1994})\BibitemShut {NoStop}%
\bibitem [{\citenamefont {Diniz}\ \emph {et~al.}(2025)\citenamefont {Diniz}, \citenamefont {Villas-Boas},\ and\ \citenamefont {Santos}}]{Diniz_2025_reset}%
  \BibitemOpen
  \bibfield  {author} {\bibinfo {author} {\bibfnamefont {C.~M.}\ \bibnamefont {Diniz}}, \bibinfo {author} {\bibfnamefont {C.~J.}\ \bibnamefont {Villas-Boas}},\ and\ \bibinfo {author} {\bibfnamefont {A.~C.}\ \bibnamefont {Santos}},\ }\href {https://doi.org/10.1088/2058-9565/adbded} {\bibfield  {journal} {\bibinfo  {journal} {Quantum Science and Technology}\ }\textbf {\bibinfo {volume} {10}},\ \bibinfo {pages} {025039} (\bibinfo {year} {2025})}\BibitemShut {NoStop}%
\bibitem [{\citenamefont {Eisberg}\ and\ \citenamefont {Resnick}(1985)}]{eisberg1985}%
  \BibitemOpen
  \bibfield  {author} {\bibinfo {author} {\bibfnamefont {R.}~\bibnamefont {Eisberg}}\ and\ \bibinfo {author} {\bibfnamefont {R.}~\bibnamefont {Resnick}},\ }\href@noop {} {\emph {\bibinfo {title} {Quantum Physics of Atoms, Molecules, Solids, Nuclei, and Particles}}},\ \bibinfo {edition} {2nd}\ ed.\ (\bibinfo  {publisher} {Wiley},\ \bibinfo {year} {1985})\BibitemShut {NoStop}%
\bibitem [{\citenamefont {Madau}\ and\ \citenamefont {Dickinson}(2014)}]{MadauDickinson2014}%
  \BibitemOpen
  \bibfield  {author} {\bibinfo {author} {\bibfnamefont {P.}~\bibnamefont {Madau}}\ and\ \bibinfo {author} {\bibfnamefont {M.}~\bibnamefont {Dickinson}},\ }\href {https://doi.org/10.1146/annurev-astro-081811-125615} {\bibfield  {journal} {\bibinfo  {journal} {Annual Review of Astronomy and Astrophysics}\ }\textbf {\bibinfo {volume} {52}},\ \bibinfo {pages} {415} (\bibinfo {year} {2014})}\BibitemShut {NoStop}%
\bibitem [{\citenamefont {Fermi}(1932)}]{fermi1932}%
  \BibitemOpen
  \bibfield  {author} {\bibinfo {author} {\bibfnamefont {E.}~\bibnamefont {Fermi}},\ }\href {https://doi.org/10.1103/RevModPhys.4.87} {\bibfield  {journal} {\bibinfo  {journal} {Reviews of Modern Physics}\ }\textbf {\bibinfo {volume} {4}},\ \bibinfo {pages} {87} (\bibinfo {year} {1932})}\BibitemShut {NoStop}%
\bibitem [{\citenamefont {Cohen-Tannoudji}\ \emph {et~al.}(1977)\citenamefont {Cohen-Tannoudji}, \citenamefont {Diu},\ and\ \citenamefont {Laloë}}]{cohen1977}%
  \BibitemOpen
  \bibfield  {author} {\bibinfo {author} {\bibfnamefont {C.}~\bibnamefont {Cohen-Tannoudji}}, \bibinfo {author} {\bibfnamefont {B.}~\bibnamefont {Diu}},\ and\ \bibinfo {author} {\bibfnamefont {F.}~\bibnamefont {Laloë}},\ }\href@noop {} {\emph {\bibinfo {title} {Quantum Mechanics, Volume 2}}}\ (\bibinfo  {publisher} {Wiley},\ \bibinfo {year} {1977})\BibitemShut {NoStop}%
\bibitem [{\citenamefont {de~Toledo~Piza}(2023)}]{toledo_piza2023}%
  \BibitemOpen
  \bibfield  {author} {\bibinfo {author} {\bibfnamefont {A.~F.~R.}\ \bibnamefont {de~Toledo~Piza}},\ }\href@noop {} {\emph {\bibinfo {title} {Mecânica Quântica}}},\ \bibinfo {edition} {2nd}\ ed.\ (\bibinfo  {publisher} {EDUSP},\ \bibinfo {address} {São Paulo},\ \bibinfo {year} {2023})\ p.\ \bibinfo {pages} {632}\BibitemShut {NoStop}%
\bibitem [{\citenamefont {Souza}\ \emph {et~al.}(2015)\citenamefont {Souza}, \citenamefont {Cabral}, \citenamefont {Oliveira},\ and\ \citenamefont {Villas-Boas}}]{Souza2015}%
  \BibitemOpen
  \bibfield  {author} {\bibinfo {author} {\bibfnamefont {J.~A.}\ \bibnamefont {Souza}}, \bibinfo {author} {\bibfnamefont {L.}~\bibnamefont {Cabral}}, \bibinfo {author} {\bibfnamefont {R.~R.}\ \bibnamefont {Oliveira}},\ and\ \bibinfo {author} {\bibfnamefont {C.~J.}\ \bibnamefont {Villas-Boas}},\ }\href {https://doi.org/10.1103/PhysRevA.92.023818} {\bibfield  {journal} {\bibinfo  {journal} {Phys. Rev. A}\ }\textbf {\bibinfo {volume} {92}},\ \bibinfo {pages} {023818} (\bibinfo {year} {2015})}\BibitemShut {NoStop}%
\bibitem [{Note1()}]{Note1}%
  \BibitemOpen
  \bibinfo {note} {Classical analogs as the symmetric and antisymmetric motions in coupled pendulum systems, familiar from undergraduate physics. In contrast, at low excitation levels, certain collective states exhibit genuinely nonclassical behavior that cannot be captured by classical theories~\cite {VillasBoas2025}.}\BibitemShut {Stop}%
\bibitem [{Note2()}]{Note2}%
  \BibitemOpen
  \bibinfo {note} {This result assumes no coupling between modes, so the collective modes remain degenerate. When matter is included, we must guarantee that we are in the regime where the matter–mode coupling is much smaller than the mode frequencies, which justifies the rotating wave approximation (RWA). For stronger couplings (e.g., in the Rabi model), modes may become non-degenerate, and the properties of the collective modes change completely.}\BibitemShut {Stop}%
\bibitem [{\citenamefont {Breuer}\ and\ \citenamefont {Petruccione}(2002)}]{breuer2002}%
  \BibitemOpen
  \bibfield  {author} {\bibinfo {author} {\bibfnamefont {H.-P.}\ \bibnamefont {Breuer}}\ and\ \bibinfo {author} {\bibfnamefont {F.}~\bibnamefont {Petruccione}},\ }\href@noop {} {\emph {\bibinfo {title} {The Theory of Open Quantum Systems}}}\ (\bibinfo  {publisher} {Oxford University Press},\ \bibinfo {address} {Oxford},\ \bibinfo {year} {2002})\BibitemShut {NoStop}%
\bibitem [{\citenamefont {Gardiner}\ and\ \citenamefont {Zoller}(2004)}]{gardiner2004}%
  \BibitemOpen
  \bibfield  {author} {\bibinfo {author} {\bibfnamefont {C.~W.}\ \bibnamefont {Gardiner}}\ and\ \bibinfo {author} {\bibfnamefont {P.}~\bibnamefont {Zoller}},\ }\href@noop {} {\emph {\bibinfo {title} {Quantum Noise: A Handbook of Markovian and Non-Markovian Quantum Stochastic Methods with Applications to Quantum Optics}}},\ \bibinfo {edition} {3rd}\ ed.\ (\bibinfo  {publisher} {Springer},\ \bibinfo {year} {2004})\BibitemShut {NoStop}%
\bibitem [{Note3()}]{Note3}%
  \BibitemOpen
  \bibinfo {note} {Should the modes interact with distinct material regions -- spaced by several wavelengths -- alternative symmetric states may arise, leading to spatially nonuniform energy transfer. Here, we are also neglecting the temperature effects of the matter since we are interested in the amount of energy of the radiation field that can be absorbed by the matter only.}\BibitemShut {Stop}%
\bibitem [{\citenamefont {Werlang}\ \emph {et~al.}(2008)\citenamefont {Werlang}, \citenamefont {Guzm\'an}, \citenamefont {Prado},\ and\ \citenamefont {Villas-B\^oas}}]{Werlang2008}%
  \BibitemOpen
  \bibfield  {author} {\bibinfo {author} {\bibfnamefont {T.}~\bibnamefont {Werlang}}, \bibinfo {author} {\bibfnamefont {R.}~\bibnamefont {Guzm\'an}}, \bibinfo {author} {\bibfnamefont {F.~O.}\ \bibnamefont {Prado}},\ and\ \bibinfo {author} {\bibfnamefont {C.~J.}\ \bibnamefont {Villas-B\^oas}},\ }\href {https://doi.org/10.1103/PhysRevA.78.033820} {\bibfield  {journal} {\bibinfo  {journal} {Phys. Rev. A}\ }\textbf {\bibinfo {volume} {78}},\ \bibinfo {pages} {033820} (\bibinfo {year} {2008})}\BibitemShut {NoStop}%
\bibitem [{\citenamefont {Prado}\ \emph {et~al.}(2009)\citenamefont {Prado}, \citenamefont {Duzzioni}, \citenamefont {Moussa}, \citenamefont {de~Almeida},\ and\ \citenamefont {Villas-B\^oas}}]{prado2009}%
  \BibitemOpen
  \bibfield  {author} {\bibinfo {author} {\bibfnamefont {F.~O.}\ \bibnamefont {Prado}}, \bibinfo {author} {\bibfnamefont {E.~I.}\ \bibnamefont {Duzzioni}}, \bibinfo {author} {\bibfnamefont {M.~H.~Y.}\ \bibnamefont {Moussa}}, \bibinfo {author} {\bibfnamefont {N.~G.}\ \bibnamefont {de~Almeida}},\ and\ \bibinfo {author} {\bibfnamefont {C.~J.}\ \bibnamefont {Villas-B\^oas}},\ }\href {https://doi.org/10.1103/PhysRevLett.102.073008} {\bibfield  {journal} {\bibinfo  {journal} {Phys. Rev. Lett.}\ }\textbf {\bibinfo {volume} {102}},\ \bibinfo {pages} {073008} (\bibinfo {year} {2009})}\BibitemShut {NoStop}%
\bibitem [{\citenamefont {Prado}\ \emph {et~al.}(2011)\citenamefont {Prado}, \citenamefont {de~Almeida}, \citenamefont {Duzzioni}, \citenamefont {Moussa},\ and\ \citenamefont {Villas-Boas}}]{prado2011}%
  \BibitemOpen
  \bibfield  {author} {\bibinfo {author} {\bibfnamefont {F.}~\bibnamefont {Prado}}, \bibinfo {author} {\bibfnamefont {N.}~\bibnamefont {de~Almeida}}, \bibinfo {author} {\bibfnamefont {E.}~\bibnamefont {Duzzioni}}, \bibinfo {author} {\bibfnamefont {M.}~\bibnamefont {Moussa}},\ and\ \bibinfo {author} {\bibfnamefont {C.}~\bibnamefont {Villas-Boas}},\ }\href@noop {} {\bibfield  {journal} {\bibinfo  {journal} {Physical Review A}\ }\textbf {\bibinfo {volume} {84}},\ \bibinfo {pages} {012112} (\bibinfo {year} {2011})}\BibitemShut {NoStop}%
\bibitem [{\citenamefont {Binder}\ \emph {et~al.}(2024)\citenamefont {Binder}, \citenamefont {Ahmed}, \citenamefont {Ayin}, \citenamefont {Barkoutsos}, \citenamefont {Berec}, \citenamefont {Elvira}, \citenamefont {Gao}, \citenamefont {Garcia}, \citenamefont {Ghosh}, \citenamefont {Granade}, \citenamefont {Johansson}, \citenamefont {Kumar}, \citenamefont {Minev}, \citenamefont {Nation}, \citenamefont {Rajan}, \citenamefont {Singh}, \citenamefont {Tantivasadakarn}, \citenamefont {Vaidya},\ and\ \citenamefont {Willsch}}]{qutip5}%
  \BibitemOpen
  \bibfield  {author} {\bibinfo {author} {\bibfnamefont {F.~C.}\ \bibnamefont {Binder}}, \bibinfo {author} {\bibfnamefont {A.}~\bibnamefont {Ahmed}}, \bibinfo {author} {\bibfnamefont {D.~A.}\ \bibnamefont {Ayin}}, \bibinfo {author} {\bibfnamefont {P.~K.}\ \bibnamefont {Barkoutsos}}, \bibinfo {author} {\bibfnamefont {V.}~\bibnamefont {Berec}}, \bibinfo {author} {\bibfnamefont {V.}~\bibnamefont {Elvira}}, \bibinfo {author} {\bibfnamefont {J.}~\bibnamefont {Gao}}, \bibinfo {author} {\bibfnamefont {S.~E.}\ \bibnamefont {Garcia}}, \bibinfo {author} {\bibfnamefont {J.}~\bibnamefont {Ghosh}}, \bibinfo {author} {\bibfnamefont {C.}~\bibnamefont {Granade}}, \bibinfo {author} {\bibfnamefont {J.~R.}\ \bibnamefont {Johansson}}, \bibinfo {author} {\bibfnamefont {R.}~\bibnamefont {Kumar}}, \bibinfo {author} {\bibfnamefont {Z.~K.}\ \bibnamefont {Minev}}, \bibinfo {author} {\bibfnamefont {P.~D.}\ \bibnamefont {Nation}}, \bibinfo {author} {\bibfnamefont {V.~P.}\ \bibnamefont {Rajan}}, \bibinfo {author} {\bibfnamefont
  {S.}~\bibnamefont {Singh}}, \bibinfo {author} {\bibfnamefont {N.}~\bibnamefont {Tantivasadakarn}}, \bibinfo {author} {\bibfnamefont {S.}~\bibnamefont {Vaidya}},\ and\ \bibinfo {author} {\bibfnamefont {D.}~\bibnamefont {Willsch}},\ }\href {https://doi.org/10.22331/q-2024-01-11-1435} {\bibfield  {journal} {\bibinfo  {journal} {Quantum}\ }\textbf {\bibinfo {volume} {8}},\ \bibinfo {pages} {1435} (\bibinfo {year} {2024})}\BibitemShut {NoStop}%
\bibitem [{\citenamefont {Farrera}\ \emph {et~al.}(2020)\citenamefont {Farrera}, \citenamefont {Brekenfeld}, \citenamefont {Niemietz}, \citenamefont {Christesen},\ and\ \citenamefont {Rempe}}]{farrera2020}%
  \BibitemOpen
  \bibfield  {author} {\bibinfo {author} {\bibfnamefont {P.}~\bibnamefont {Farrera}}, \bibinfo {author} {\bibfnamefont {M.}~\bibnamefont {Brekenfeld}}, \bibinfo {author} {\bibfnamefont {D.}~\bibnamefont {Niemietz}}, \bibinfo {author} {\bibfnamefont {J.~D.}\ \bibnamefont {Christesen}},\ and\ \bibinfo {author} {\bibfnamefont {G.}~\bibnamefont {Rempe}},\ }\href@noop {} {\bibinfo {title} {Single trapped atoms coupled to crossed fiber-cavities}},\ \bibinfo {howpublished} {\url{https://premc.org/doc/QTech2020/2_Pau_Farrera_Slides.pdf}} (\bibinfo {year} {2020}),\ \bibinfo {note} {presented at QTech 2020}\BibitemShut {NoStop}%
\bibitem [{\citenamefont {Hohmann}(2023)}]{hohmann2023}%
  \BibitemOpen
  \bibfield  {author} {\bibinfo {author} {\bibfnamefont {C.}~\bibnamefont {Hohmann}},\ }\href {https://doi.org/10.1002/phvs.202200015} {\bibfield  {journal} {\bibinfo  {journal} {Physik in unserer Zeit}\ }\textbf {\bibinfo {volume} {54}},\ \bibinfo {pages} {4} (\bibinfo {year} {2023})}\BibitemShut {NoStop}%
\bibitem [{\citenamefont {Solak}\ \emph {et~al.}(2024{\natexlab{a}})\citenamefont {Solak}, \citenamefont {Rossatto},\ and\ \citenamefont {Villas-Boas}}]{Solak2024}%
  \BibitemOpen
  \bibfield  {author} {\bibinfo {author} {\bibfnamefont {L.~O.~R.}\ \bibnamefont {Solak}}, \bibinfo {author} {\bibfnamefont {D.~Z.}\ \bibnamefont {Rossatto}},\ and\ \bibinfo {author} {\bibfnamefont {C.~J.}\ \bibnamefont {Villas-Boas}},\ }\href {https://doi.org/10.1103/PhysRevA.109.062620} {\bibfield  {journal} {\bibinfo  {journal} {Phys. Rev. A}\ }\textbf {\bibinfo {volume} {109}},\ \bibinfo {pages} {062620} (\bibinfo {year} {2024}{\natexlab{a}})}\BibitemShut {NoStop}%
\bibitem [{\citenamefont {Solak}\ \emph {et~al.}(2024{\natexlab{b}})\citenamefont {Solak}, \citenamefont {Villas-Boas},\ and\ \citenamefont {Rossatto}}]{solak2024beamsplitter}%
  \BibitemOpen
  \bibfield  {author} {\bibinfo {author} {\bibfnamefont {L.~O.~R.}\ \bibnamefont {Solak}}, \bibinfo {author} {\bibfnamefont {C.~J.}\ \bibnamefont {Villas-Boas}},\ and\ \bibinfo {author} {\bibfnamefont {D.~Z.}\ \bibnamefont {Rossatto}},\ }\href {https://arxiv.org/abs/2408.15059} {\bibinfo {title} {Beam splitter for dark and bright states of light}} (\bibinfo {year} {2024}{\natexlab{b}}),\ \Eprint {https://arxiv.org/abs/2408.15059} {arXiv:2408.15059 [quant-ph]} \BibitemShut {NoStop}%
\bibitem [{\citenamefont {Seubert}\ \emph {et~al.}(2025)\citenamefont {Seubert}, \citenamefont {Hartung}, \citenamefont {Welte}, \citenamefont {Rempe},\ and\ \citenamefont {Distante}}]{seubert2025}%
  \BibitemOpen
  \bibfield  {author} {\bibinfo {author} {\bibfnamefont {M.}~\bibnamefont {Seubert}}, \bibinfo {author} {\bibfnamefont {L.}~\bibnamefont {Hartung}}, \bibinfo {author} {\bibfnamefont {S.}~\bibnamefont {Welte}}, \bibinfo {author} {\bibfnamefont {G.}~\bibnamefont {Rempe}},\ and\ \bibinfo {author} {\bibfnamefont {E.}~\bibnamefont {Distante}},\ }\href {https://doi.org/10.1103/PRXQuantum.6.010322} {\bibfield  {journal} {\bibinfo  {journal} {PRX Quantum}\ }\textbf {\bibinfo {volume} {6}},\ \bibinfo {pages} {010322} (\bibinfo {year} {2025})}\BibitemShut {NoStop}%
\bibitem [{\citenamefont {Aghanim}\ \emph {et~al.}(2020)\citenamefont {Aghanim}, \citenamefont {Akrami}, \citenamefont {Ashdown}, \citenamefont {Aumont}, \citenamefont {Baccigalupi}, \citenamefont {Ballardini}, \citenamefont {Banday}, \citenamefont {Barreiro}, \citenamefont {Bartolo}, \citenamefont {Basak},\ and\ \citenamefont {et~al.}}]{planck2018}%
  \BibitemOpen
  \bibfield  {author} {\bibinfo {author} {\bibfnamefont {N.}~\bibnamefont {Aghanim}}, \bibinfo {author} {\bibfnamefont {Y.}~\bibnamefont {Akrami}}, \bibinfo {author} {\bibfnamefont {M.}~\bibnamefont {Ashdown}}, \bibinfo {author} {\bibfnamefont {J.}~\bibnamefont {Aumont}}, \bibinfo {author} {\bibfnamefont {C.}~\bibnamefont {Baccigalupi}}, \bibinfo {author} {\bibfnamefont {M.}~\bibnamefont {Ballardini}}, \bibinfo {author} {\bibfnamefont {A.~J.}\ \bibnamefont {Banday}}, \bibinfo {author} {\bibfnamefont {R.~B.}\ \bibnamefont {Barreiro}}, \bibinfo {author} {\bibfnamefont {N.}~\bibnamefont {Bartolo}}, \bibinfo {author} {\bibfnamefont {S.}~\bibnamefont {Basak}},\ and\ \bibinfo {author} {\bibnamefont {et~al.}},\ }\href {https://doi.org/10.1051/0004-6361/201833910} {\bibfield  {journal} {\bibinfo  {journal} {Astronomy \& Astrophysics}\ }\textbf {\bibinfo {volume} {641}},\ \bibinfo {pages} {A6} (\bibinfo {year} {2020})}\BibitemShut {NoStop}%
\bibitem [{\citenamefont {Collaboration}(2022)}]{desy3}%
  \BibitemOpen
  \bibfield  {author} {\bibinfo {author} {\bibfnamefont {D.}~\bibnamefont {Collaboration}},\ }\href@noop {} {\bibinfo {title} {Dark energy survey year 3 cosmology results}} (\bibinfo {year} {2022}),\ \bibinfo {note} {\url{https://www.darkenergysurvey.org/des-year-3-cosmology-results/}}\BibitemShut {NoStop}%
\bibitem [{\citenamefont {Copeland}\ \emph {et~al.}(2006)\citenamefont {Copeland}, \citenamefont {Sami},\ and\ \citenamefont {Tsujikawa}}]{copeland2006}%
  \BibitemOpen
  \bibfield  {author} {\bibinfo {author} {\bibfnamefont {E.~J.}\ \bibnamefont {Copeland}}, \bibinfo {author} {\bibfnamefont {M.}~\bibnamefont {Sami}},\ and\ \bibinfo {author} {\bibfnamefont {S.}~\bibnamefont {Tsujikawa}},\ }\href {https://doi.org/10.1142/S021827180600942X} {\bibfield  {journal} {\bibinfo  {journal} {International Journal of Modern Physics D}\ }\textbf {\bibinfo {volume} {15}},\ \bibinfo {pages} {1753–1936} (\bibinfo {year} {2006})}\BibitemShut {NoStop}%
\bibitem [{\citenamefont {Amendola}\ and\ \citenamefont {Tsujikawa}(2010)}]{amendola2010}%
  \BibitemOpen
  \bibfield  {author} {\bibinfo {author} {\bibfnamefont {L.}~\bibnamefont {Amendola}}\ and\ \bibinfo {author} {\bibfnamefont {S.}~\bibnamefont {Tsujikawa}},\ }\href@noop {} {\emph {\bibinfo {title} {Dark Energy: Theory and Observations}}}\ (\bibinfo  {publisher} {Cambridge University Press},\ \bibinfo {year} {2010})\BibitemShut {NoStop}%
\bibitem [{\citenamefont {Frieman}\ \emph {et~al.}(2008)\citenamefont {Frieman}, \citenamefont {Turner},\ and\ \citenamefont {Huterer}}]{frieman2008}%
  \BibitemOpen
  \bibfield  {author} {\bibinfo {author} {\bibfnamefont {J.~A.}\ \bibnamefont {Frieman}}, \bibinfo {author} {\bibfnamefont {M.~S.}\ \bibnamefont {Turner}},\ and\ \bibinfo {author} {\bibfnamefont {D.}~\bibnamefont {Huterer}},\ }\href {https://doi.org/10.1146/annurev.astro.46.060407.145243} {\bibfield  {journal} {\bibinfo  {journal} {Annual Review of Astronomy and Astrophysics}\ }\textbf {\bibinfo {volume} {46}},\ \bibinfo {pages} {385–432} (\bibinfo {year} {2008})}\BibitemShut {NoStop}%
\bibitem [{\citenamefont {Zwicky}(1933)}]{zwicky1933}%
  \BibitemOpen
  \bibfield  {author} {\bibinfo {author} {\bibfnamefont {F.}~\bibnamefont {Zwicky}},\ }\href@noop {} {\bibfield  {journal} {\bibinfo  {journal} {Helvetica Physica Acta}\ }\textbf {\bibinfo {volume} {6}},\ \bibinfo {pages} {110} (\bibinfo {year} {1933})}\BibitemShut {NoStop}%
\bibitem [{\citenamefont {Bertone}\ \emph {et~al.}(2005)\citenamefont {Bertone}, \citenamefont {Hooper},\ and\ \citenamefont {Silk}}]{bertone2005}%
  \BibitemOpen
  \bibfield  {author} {\bibinfo {author} {\bibfnamefont {G.}~\bibnamefont {Bertone}}, \bibinfo {author} {\bibfnamefont {D.}~\bibnamefont {Hooper}},\ and\ \bibinfo {author} {\bibfnamefont {J.}~\bibnamefont {Silk}},\ }\href {https://doi.org/10.1016/j.physrep.2004.08.031} {\bibfield  {journal} {\bibinfo  {journal} {Physics Reports}\ }\textbf {\bibinfo {volume} {405}},\ \bibinfo {pages} {279} (\bibinfo {year} {2005})}\BibitemShut {NoStop}%
\bibitem [{\citenamefont {Collaboration}\ \emph {et~al.}(2020)\citenamefont {Collaboration}, \citenamefont {Aghanim}, \citenamefont {Akrami} \emph {et~al.}}]{planck2020}%
  \BibitemOpen
  \bibfield  {author} {\bibinfo {author} {\bibfnamefont {P.}~\bibnamefont {Collaboration}}, \bibinfo {author} {\bibfnamefont {N.}~\bibnamefont {Aghanim}}, \bibinfo {author} {\bibfnamefont {Y.}~\bibnamefont {Akrami}}, \emph {et~al.},\ }\href {https://doi.org/10.1051/0004-6361/201833910} {\bibfield  {journal} {\bibinfo  {journal} {Astronomy \& Astrophysics}\ }\textbf {\bibinfo {volume} {641}},\ \bibinfo {pages} {A6} (\bibinfo {year} {2020})}\BibitemShut {NoStop}%
\bibitem [{\citenamefont {Shull}\ \emph {et~al.}(2012)\citenamefont {Shull}, \citenamefont {Smith},\ and\ \citenamefont {Danforth}}]{shull2012}%
  \BibitemOpen
  \bibfield  {author} {\bibinfo {author} {\bibfnamefont {J.~M.}\ \bibnamefont {Shull}}, \bibinfo {author} {\bibfnamefont {B.~D.}\ \bibnamefont {Smith}},\ and\ \bibinfo {author} {\bibfnamefont {C.~W.}\ \bibnamefont {Danforth}},\ }\href {https://doi.org/10.1088/0004-637X/759/1/23} {\bibfield  {journal} {\bibinfo  {journal} {The Astrophysical Journal}\ }\textbf {\bibinfo {volume} {759}},\ \bibinfo {pages} {23} (\bibinfo {year} {2012})}\BibitemShut {NoStop}%
\bibitem [{\citenamefont {Parke}\ \emph {et~al.}(2024)\citenamefont {Parke}, \citenamefont {Thomm}, \citenamefont {Santos}, \citenamefont {Cidrim}, \citenamefont {Higgins}, \citenamefont {Mallweger}, \citenamefont {Kuk}, \citenamefont {Salim}, \citenamefont {Bachelard}, \citenamefont {Villas-Boas} \emph {et~al.}}]{parke2024phononic}%
  \BibitemOpen
  \bibfield  {author} {\bibinfo {author} {\bibfnamefont {H.}~\bibnamefont {Parke}}, \bibinfo {author} {\bibfnamefont {R.}~\bibnamefont {Thomm}}, \bibinfo {author} {\bibfnamefont {A.~C.}\ \bibnamefont {Santos}}, \bibinfo {author} {\bibfnamefont {A.}~\bibnamefont {Cidrim}}, \bibinfo {author} {\bibfnamefont {G.}~\bibnamefont {Higgins}}, \bibinfo {author} {\bibfnamefont {M.}~\bibnamefont {Mallweger}}, \bibinfo {author} {\bibfnamefont {N.}~\bibnamefont {Kuk}}, \bibinfo {author} {\bibfnamefont {S.}~\bibnamefont {Salim}}, \bibinfo {author} {\bibfnamefont {R.}~\bibnamefont {Bachelard}}, \bibinfo {author} {\bibfnamefont {C.~J.}\ \bibnamefont {Villas-Boas}}, \emph {et~al.},\ }\href@noop {} {\bibfield  {journal} {\bibinfo  {journal} {arXiv preprint arXiv:2403.07154}\ } (\bibinfo {year} {2024})}\BibitemShut {NoStop}%
\bibitem [{\citenamefont {Schönberg}(1952)}]{schonberg1952}%
  \BibitemOpen
  \bibfield  {author} {\bibinfo {author} {\bibfnamefont {M.}~\bibnamefont {Schönberg}},\ }\href@noop {} {\bibfield  {journal} {\bibinfo  {journal} {Il Nuovo Cimento}\ }\textbf {\bibinfo {volume} {9}},\ \bibinfo {pages} {1139} (\bibinfo {year} {1952})}\BibitemShut {NoStop}%
\bibitem [{\citenamefont {Schönberg}(1953{\natexlab{a}})}]{schonberg1953a}%
  \BibitemOpen
  \bibfield  {author} {\bibinfo {author} {\bibfnamefont {M.}~\bibnamefont {Schönberg}},\ }\href@noop {} {\bibfield  {journal} {\bibinfo  {journal} {Il Nuovo Cimento}\ }\textbf {\bibinfo {volume} {10}},\ \bibinfo {pages} {419} (\bibinfo {year} {1953}{\natexlab{a}})}\BibitemShut {NoStop}%
\bibitem [{\citenamefont {Schönberg}(1953{\natexlab{b}})}]{schonberg1953b}%
  \BibitemOpen
  \bibfield  {author} {\bibinfo {author} {\bibfnamefont {M.}~\bibnamefont {Schönberg}},\ }\href@noop {} {\bibfield  {journal} {\bibinfo  {journal} {Il Nuovo Cimento}\ }\textbf {\bibinfo {volume} {10}},\ \bibinfo {pages} {697} (\bibinfo {year} {1953}{\natexlab{b}})}\BibitemShut {NoStop}%
\end{thebibliography}%

\newpage

\appendix


\onecolumngrid
\newpage

\begin{center}
	{\large{ {\bf Supplemental Material for: \\ Dark States of Light and the Hidden Energy in Thermal Radiation Detection}}}

\vskip0.5\baselineskip{Celso Jorge Villas-Boas~\orcidlink{0000-0001-5622-786X}$^{1}$, and Ciro Micheletti~Diniz~\orcidlink{0000-0002-7602-0468}$^{1}$}

\vskip0.5\baselineskip{

   {\em $^{1}${{Departamento de Física, Universidade Federal de São Carlos, Rodovia Washington Luís, km 235 - SP-310, 13565-905 São Carlos, SP, Brazil}}}
}


\end{center}

\setcounter{equation}{0}
\setcounter{figure}{0}
\setcounter{table}{0}

\renewcommand{\theequation}{S\arabic{equation}}
\renewcommand{\thefigure}{S\arabic{figure}}
\renewcommand{\bibnumfmt}[1]{[S#1]}
\renewcommand{\citenumfont}[1]{S#1}


\section*{Number of collective bright and dark states for a phase-fixed change-of-basis matrix}

We can build the collective operators as a function of the lowering operators $\hat{a}_j$ considering the Fock basis. Thus, we can write the collective operators as~\cite{VillasBoas2025}
\begin{equation}
    \hat{A}_\mu = \sum^{M}_{j=1} U_{\mu j}\hat{a}_j,
\end{equation}
\noindent where $1 \leq \mu \leq M$ and $U$ is the change-of-basis matrix.  When the number of modes $M$ is even, we can construct $U$ making $O_{1,j}=1/\sqrt{M}$, $O_{2,j}=(-1)^{j-1}/\sqrt{M}$, and the other rows are obtained by the combinations of $\pm 1/\sqrt{M}$ that satisfy the linear independence of the rows. When $M$ is odd, the construction of the change-of-basis matrix is a bit trickier. However, the first row still is $O_{1,j}=1/\sqrt{M}$, and now the other rows are constructed in order to achieve an orthonormal set of operators. The orthogonality between the operators ensures the canonical commutation relations between them, while the normalization conserves the total number of excitations.

Since the ground state is the same in both bases, the eigenstates of the collective basis can be written as
\begin{equation}\label{ket colec basis}
    \ket{\Psi^N_{n_0, n_1, \ldots, n_{M-1}}} = \ket{n_0, n_1, \ldots, n_{M-2}, N - n_0 - \ldots - n_{M-2}}_{0', 1', \ldots, (M-1)'} = \prod^{M-1}_{\mu=0}\frac{\left( \hat{A}^\dagger_\mu \right)^{n_\mu}}{\sqrt{n_\mu !}}\ket{0,0, \ldots, 0}_{0', 1', \ldots, (M-1)'}, 
\end{equation}
\noindent where $\hat{A}_\mu^\dagger = (\hat{A}_\mu)^\dagger$ is the raising operator of the $\mu$-th collective mode (representing by the subindex $\mu'$), and we have the condition $\sum^{M-1}_{\mu=0}n_\mu = N$, and, consequently, $n_{M-1} = N - \sum^{M-2}_{\mu=0}n_\mu$. Therefore, the $\mu$-th collective mode has $n_\mu$ photons.

The first mode ($\mu=0$) is the only symmetric one and, consequently, the only collective mode that can exchange photons because it couples to the matter. For this reason, considering an arbitrary number of excitations $N$, the collective states that have all photons in the symmetric mode are named by bright states once they can exchange the possible maximum amount of energy and reads $\ket{\Psi^N_{N, 0, \ldots, 0, 0}}$. Therefore, given a number of excitations $N$ and modes $M$, there will always be a unique bright state since there is just one way to distribute $N$ indistinguishable photons in $M$ different modes in such a way that all photons can be exchanged. On the other hand, the amount of dark states depends not only on the number of modes $M$ but also on the number of excitations $N$. Considering $M$ modes and $N=1$, the set of dark states is $\left\{ \ket{\Psi^1_{0,1,0,\ldots,0,0}}, \ket{\Psi^1_{0,0,1,\ldots,0,0}}, \ldots, \ket{\Psi^1_{0,0,0,\ldots,1,0}}, \ket{\Psi^1_{0,0,0,\ldots,0,1}} \right\}$, which means there are $M-1$ dark states in this case. However, with $N>1$, this amount increases. We can easily see this increasing by considering $N=2$ and $M=3$. In this case, the dark states are $\left\{ \ket{\Psi^2_{0,2,0}}, \ket{\Psi^2_{0,0,2}}, \ket{\Psi^1_{0,1,1}} \right\}$, which means that are three dark states, one more when compared to the case with $N=1$ and $M=3$, where there are two dark states $\left\{ \ket{\Psi^2_{0,1,0}}, \ket{\Psi^2_{0,0,1}} \right\}$. Hence, for the general case of $M$ modes and $N$ excitations, the number of dark states $N_\mathrm{DS}$ can be found by solving the combinatorial problem that computes the different ways to distribute $N$ photons in $M-1$ modes. This result is also the number of possible values that each $n_\mu$ can assume while respecting the conditions $\sum^{M-1}_{\mu=0}n_\mu = N$ and $n_0 = 0$, which is given by
\begin{equation}
    N_\mathrm{DS} = \sum^{N}_{n_1=0} \sum^{N-n_1}_{n_2=0} \ldots \sum^{N-n_1 - \ldots -n_{M-4}}_{n_{M-3}=0} \sum^{N-n_1 - \ldots -n_{M-3}}_{n_{M-2}=0}  1,
\end{equation}
\noindent where we can notice the lack of the respective sum of $n_0$ because in all dark states $n_0 = 0$, and the absence of a sum with index $n_{M-1}$ because, once the number of photons in the other modes is known, the number of photons in the last mode must be the one that satisfies $\sum^{M-1}_{\mu=0}n_\mu = N$ (see the Supplemental Material of~\cite{diniz2024}). 

Similarly, the number of all the collective states $N_\mathrm{CS}$ is given by
\begin{equation}
    N_\mathrm{CS} = \sum^{N}_{n_0=0} \sum^{N-n_0}_{n_1=0} \ldots \sum^{N-n_0 - \ldots -n_{M-4}}_{n_{M-3}=0} \sum^{N-n_0 - \ldots -n_{M-3}}_{n_{M-2}=0}  1,
\end{equation}
where now we have the summation over $n_0$ because we need to consider the distributions that take into account the presence of photons in the symmetric mode. In other words, we are now interested in computing the number of states forming the collective basis that describes a problem with $M$ modes and $N$ excitations, which are bright, dark, or intermediate states~\cite{VillasBoas2025}.

\section*{Energy exchange and decay of the collective states}

When written as a function of the collective operators, the electrical operator becomes $E^{+} = \sqrt{M}\hat{A}_0$~\cite{VillasBoas2025}, depending only on the symmetric collective mode. In this sense, as discussed above, just the first mode is able to exchange energy. Thus, the interaction between the general state (in the collective basis) $\ket{\Psi^N_{n_0, n_1, \ldots, n_{M-2}, n_{M-1}}}$ and the atom results in
\begin{equation}
\begin{aligned}
     H\ket{\Psi^N_{n_0, n_1, \ldots, n_{M-2}, n_{M-1}}}\ket{g} & = gE^+\sigma_+\ket{\Psi^N_{n_0, n_q, \ldots, n_{M-2}, n_{M-1}}}\ket{g} = \\ & g\sqrt{M}\hat{A}_0\sigma_+\ket{\Psi^N_{n_0, n_1, \ldots, n_{M-2}, n_{M-1}}}\ket{g} = g\sqrt{M}\sqrt{n_0}\ket{\Psi^{N-1}_{n_0-1, n_1, \ldots, n_{M-2}, n_{M-1}}}\ket{e},
\end{aligned} 
\end{equation}
\noindent where $H = g\left(E^{(+)}\sigma_{+} + E^{(-)}\sigma_{-}\right)$ is the Jaynes-Cummings interaction Hamiltonian~\cite{Glauber1963_1}, $g$ is the coupling strength, and we have omitted the tensor product symbol for the sake of simplicity. From the above equation, we can see that one photon is exchanged between the set of modes and the atom, following what happens when we consider the Fock basis. Additionally, the photon exchanged was present in the collective symmetric mode, while the photons of the other collective (antisymmetric) modes remained unaffected.

Now, assuming a constant coupling and a dissipative atom with a decay rate, we can consider that the atom returns to its ground state after being excited. Thus, the subsequent interaction between the modes and the atom, followed by the photon dissipation, leads to the final state $\ket{\Psi^{N-2}_{n_0-2, n_1, \ldots, n_{M-2}, n_{M-1}}}\ket{g}$. After successive similar interactions, all photons in the symmetric mode will be exchanged. At this time, the system reaches the steady state, and no photons will be present in the symmetric mode. Hence, after the dynamics, $n01$ photons were exchanged and dissipated, all of them from the symmetric mode, yielding the final state $\ket{\Psi^{N-n_0}_{0, n_1, \ldots, n_{M-2}, n_{M-1}}}\ket{g}$ of the whole system ($M$ modes and one atom). As one can notice, the final collective state for the modes $\ket{\Psi^{N-{n_0}}_{0, n_1, \ldots, n_{M-2}, n_{M-1}}}$ is a dark state, once there are no photons in the symmetric mode, as expected since it is the only one that interacts with matter, and the final total number of excitations is $N-n_0$.

In this context, when the initial state is bright, we have $n_0=N$, resulting in the final state (after the photon exchanges and considering the decay of the atom) $\ket{\Psi^0_{0, 0, \ldots, 0, 0}}$, where all photons were dissipated since they were initially in the symmetric mode. Contrarily, when the initial state is dark, we have $n_0=0$, as a, by consequence, no photons can be exchanged and the state remains the same, or in other words, there are no dynamics at all. Still, in the cases when $n_0 \neq 0, N$, the initial state is an intermediate state, and the respective steady state will be $\ket{\Psi^{N-n_0}_{0, n_1, \ldots, n_{M-2}, n_{M-1}}}$.

\section*{Non-null projection of thermal states in dark states}

The general thermal density matrix written in the Fock basis is given by
\begin{subequations}\label{thermal state}
\begin{equation}\label{thermal density}
    \hat{\rho} = \sum_{n}^{\infty} P_{n} \ketbra{n}{n}; 
\end{equation}
\begin{equation}\label{thermal weigth}
    P_{n} = \left( \frac{1}{1+\overline{n}} \right) \left( \frac{\overline{n}}{1+\overline{n}} \right)^n,
\end{equation}
\end{subequations}
\noindent where the mean number of photons $\overline{n} =\overline{n}(\omega, T) = \left(e^{\frac{\hbar\omega}{k T}} - 1\right)^{-1}$ depends on the temperature $T$ and the mode frequency $\omega$, following the Bose-Einstein distribution. 

Considering a mode with such frequency and at a temperature, such that we can consider that there is one excitation at most. Thus, its thermal density matrix can be approximated by $\hat{\rho} \simeq P_1\ketbra{0}{0} + P_1\ketbra{1}{1}$, with the weight $P_i$ given by Eq.~\eqref{thermal weigth}. If we have two resonant modes as the previous one, the state reads
\begin{equation}
    \hat{\rho}_{M=2} = \hat{\rho}_1\otimes\hat{\rho}_2 \simeq P_0 ^2\ketbra{0,0}{0,0} + P_0P_1\ketbra{0,1}{0,1} + P_0P_1\ketbra{1,0}{1,0} + P_1^2\ketbra{1,1}{1,1},
\end{equation}
\noindent where the subindex $M$ represents the number of modes considered, and $\ket{i,j} = \bra{i,j}^\dagger$ is the tensor product between the state $\ket{i}$ of the first mode and the state $\ket{j}$ of the second mode. The dark state in this case, written in the Fock basis, is $\ket{\Psi^1_{0,1}} = \left( 1/\sqrt{2} \right)\left( \ket{1,0} - \ket{0,1} \right)$~\cite{VillasBoas2025}. Thus, it is straightforward to notice that the collective thermal state has a non-null projection onto this dark state. Specifically,
\begin{equation}
    \bra{\Psi^1_{0,1}}\hat{\rho}_{M=2} = \frac{P_0P_1}{\sqrt{2}}\left( \ketbra{0,1}{0,1} - \ketbra{1,0}{1,0}\right) = P_0P_1\ket{\Psi^1_{0,1}}.
\end{equation}
\noindent Therefore, some fraction of the energy stored in this collective thermal state can not be exchanged. Consequently, even if these modes are continuously interacting with a dissipative atom, as discussed previously in this material, the steady state of the modes will have an energy different from zero. However, this energy is distributed in such a way that it can not be exchanged. Because of this, if we try to measure (interact) with these modes after they reach the steady state, we will not detect anything, and the modes will seem as if they are in their ground states.

Considering now the extension to $M$ modes, the full density matrix for the identical $M$ thermal states, i.e., the average thermal number of photons $\overline{n}$ is the same for the $M$ modes, reads
\begin{equation}
    \hat{\rho}_{M} = \bigotimes_{k=1}^{M} \hat{\rho}_{k}(\omega, T), 
\end{equation}
where $\hat{\rho}_{k}$ is the thermal density matrix of the $k$-th mode, which is given by Eq.~\eqref{thermal state}. Additionally, we can write the full state as 
\begin{equation}
\begin{aligned}
    \hat{\rho}_{M} = \sum_{n_1}^{\infty}\ldots\sum_{n_M}^{\infty} \left( \frac{1}{1+\overline{n}} \right)^M \left( \frac{\overline{n}}{1+\overline{n}} \right)^{n_1 + \ldots + n_{M}}\ketbra{n_1,\ldots,n_M}{n_1,\ldots,n_M},  
\end{aligned}
\end{equation}
\noindent where $n_j$ is the number of photons in the $j$-th mode, corresponding to the Fock state $\ket{n_j}$.

Considering $n_1 + n_2 + \ldots + n_M = N$, where $N$ is the total number of excitations associated with the $M$ modes, we can regroup the sums above in such a way to rewrite them as a sum for $N$ and parameterize the other sums as a function of $N$. Hence, the previous equation is rewritten as
\begin{equation}
\hat{\rho}_{M} = \sum_{N=0}^{\infty} \left(\frac{\overline{n}^N}{\left(1+\overline{n}\right)^{N+M}}\right)\sum_{n_{0}=0}^{N}\ldots\sum_{n_{M-2}=0}^{N - n_{0} \ldots - n_{M-3}}\ketbra{n_0,\ldots,N - n_{0} \ldots - n_{M-1}}{n_0,\ldots,N - n_{0} \ldots - n_{M-2}}.
\end{equation}

Each $N$ in the previous equation leads to a different set of states $\ket{n_0, \ldots, N-n_0\ldots -n_{M-1}}$. However, for each $N$, the coefficient multiplying each of these states forming the total matrix is the same. Therefore, we can express the total density matrix as $\hat{\rho}_{M} = \sum_{N=0}^{\infty} \left(\frac{\overline{n}^N}{\left(1+\overline{n}\right)^{N+M}}\right)\tilde{\hat{\rho}}$, where the non-normalized density matrix $\tilde{\rho}$ reads
\begin{equation}
    \tilde{\hat{\rho}} = \sum_{n_{0}=0}^{N}\ldots\sum_{n_{M-2}=0}^{N - n_{0} \ldots - n_{M-3}}\ketbra{n_0,\ldots,N - n_{0} \ldots - n_{M-2}}{n_0,\ldots,N - n_{0} \ldots - n_{M-2}},
\end{equation}
and it is equal to the identity matrix of the corresponding dimension. Thus, since this matrix is the identity, we can apply any transformation without changing the state. In particular, we can change from the Fock basis to the collective basis. Since it is the identity matrix, we just change how we represent the state in order to indicate that we are on a different basis. In this sense, the total thermal density matrix, in the collective basis, now reads
\begin{equation}
\hat{\rho}_{M} = \sum_{N=0}^{\infty} \left(\frac{\overline{n}^N}{\left(1+\overline{n}\right)^{N+M}}\right)\sum_{n_{0}=0}^{N}\ldots\sum_{n_{M-2}=0}^{N - n_{0} \ldots - n_{M-3}}\ketbra{\Psi_{n_0,\ldots,n_{M-2},n_{M-1}}^N}{\Psi_{n_0,\ldots,n_{M-2}, n_{M-1}}^N}.
\end{equation}
\noindent From this result, it is clear that the collective thermal state has non-null projections onto the dark states, regardless of the number of resonant modes (since $M>1$, obviously).

\section*{Initial and final number of photons present in collective thermal states}

From construction, each $\ket{\Psi_{n_0,\ldots,n_{M-2}, n_{M-1}}^N}$ or $\ket{n_0,\ldots,N-n_0\ldots-n_{M-2}}$ has $N$ photons. Still, as stated before, for a total number of excitations $N$ fixed, the coefficient multiplying the terms of the non-normalized matrix is the same. Thus, since we showed that thermal states have a projection onto dark states, looking at the total number of excitations before and after the dynamics for a set of collective states, considering a fixed $N$ can bring some light and help to better understand the problem. Hence, initially, the number of excitations is given by $N^{(\mathrm{i})}_{\mathrm{full}}=NQ_S(N,M)$, where the superindex $\mathrm{i}$ indicates that this quantity is computed at the beginning of the dynamics, $N$ is the number of photons present in each state, and $Q_S(N,M)$ represents the quantity of collectives states for $M$ modes that have $N$ photons, computed as
\begin{equation}
    Q_S(N,M) = \sum_{n_{0}=0}^{N} \sum_{n_{1}=0}^{N-n_0}\ldots\sum_{n_{M-3}=0}^{N - n_{0} \ldots - n_{M-4}}\sum_{n_{M-2}=0}^{N - n_{0} \ldots - n_{M-4} - n_{M-3}} 1.
\end{equation}
\noindent Again, the absence of a sum with index $n_{M-1}$ because, once the number of photons in the other modes is known, the number of photons in the last mode must be the one that satisfies $\sum^{M-1}_{\mu=0}n_\mu = N$. Additionally, following a similar argument, we can find the amount of existing bright ($n_0 = N$), dark ($n_0=0$), and intermediate ($n_0 \neq 0,N$) states. To do so, we need to notice that the quantity of states that have $n_0$ photons in the symmetric collective mode is given by 
\begin{equation}
    Q^{(n_0)}_S(N,M) = \sum_{n_{1}=0}^{N-n_0}\ldots\sum_{n_{M-3}=0}^{N - n_{0} \ldots - n_{M-4}}\sum_{n_{M-2}=0}^{N - n_{0} \ldots - n_{M-4} - n_{M-3}} 1,
\end{equation}
\noindent being $Q^{(n_0)}_S(N,M)$ also a function of $n_0$. Consequently, the collective states that have $n_0$ photons in the symmetric mode have together a total number of excitations given by $NQ^{(n_0)}_S(N,M)$. Equivalently, one can compute the number of states (and total excitations) considering any $n_\mu$ photons in the $\mu$-th collective mode. For didactic purposes, we choose to present the case for the symmetric mode because, with this result in hand, we can calculate the total number of excitations in a situation where all the photons from the symmetric mode are exchanged and dissipated.

In order to compute the final number of photons, we need to remember that, in the steady state, this is, after successive interactions between the modes and the dissipative atom, all the photons that were present in the symmetric mode are dissipated. Hence, it is immediate that the number of photons dissipated from the state $\ket{\Psi_{n_0,\ldots,n_{M-2}, n_{M-1}}^N}$ is $n_0$. Therefore, the number of photons present in this state at the end of the dynamics is $N-n_0$. Employing the previous result, where we calculate the number of collective states that have $n_0$ photons in the symmetric mode as $Q^{(n_0)}_S(N,M)$, we can find the final number of excitations. However, in this case, the number of excitations is not $N$, but $N-n_0$, because $n_0$ excitations were dissipated. Hence, adding up in $n_0$ yields 
\begin{equation}
    N^{(\mathrm{f})}_{\mathrm{full}} = \sum^{N}_{n_0=0} \left(N-n_0\right) Q^{(n_0)}_S(N,M) = \sum^{N}_{n_0=0} \left(N-n_0\right)\sum_{n_{1}=0}^{N-n_0}\ldots\sum_{n_{M-3}=0}^{N - n_{0} \ldots - n_{M-4}}\sum_{n_{M-2}=0}^{N - n_{01} \ldots - n_{M-4} - n_{M-3}} 1,
\end{equation}
\noindent where the superindex $\mathrm{f}$ indicates that this quantity is computed at the end of the dynamics. This result says that the number of states that had $n_0$ photons before the interaction with the dissipative atom is $Q^{(n_0)}_S(N,M)$, and these states have at the end of the dynamics $N-n_0$ photons. When we consider all the possible values for $n_0$, we achieve the total number of excitations after the dynamics for a set of collective states considering a fixed total number of excitations $N$ in the $M$ modes. 

Since we are dealing with incoherent photons, this problem resembles the combinatorial problem of computing the possible ways of distributing \textbf{N} indistinguishable objects (photons) in \textbf{M} different shelves (modes). Hence, following the same logic from the combinatorial problem, we have~\cite{andrews1998theory}
\begin{equation}
    Q^{(n_0)}_S(N,M) = \sum_{n_{1}=0}^{N-n_0}\ldots\sum_{n_{M-3}=0}^{N - n_{0} \ldots - n_{M-4}}\sum_{n_{M-2}=0}^{N - n_{0} \ldots - n_{M-4} - n_{M-3}} 1 = \binom{N-n_0+M-2}{M-2},
\end{equation}
\noindent where now we have $N-n_0$ photons to be distributed in $M-1$ modes while preserving the condition $\sum^{M-1}_{\mu=0}n_\mu = N$. Similarly, for the initial condition, we have the description but with a total of $N$ photons in such a way that the ratio between the final and initial number of excitations is
\begin{equation}\label{ratio sums}
    N^{(\mathrm{f})}_{\mathrm{full}}/N^{(\mathrm{i})}_{\mathrm{f}} = \frac{\sum^{N}_{n_0=0}(N-n_0)\binom{N-n_0+M-2}{M-2}}{\sum^{N}_{n_0=0}N\binom{N-n_0+M-2}{M-2}}.
\end{equation}
This approach facilities the following calculation of the ratio $N^{(\mathrm{f})}_{\mathrm{full}}/N^{(\mathrm{i})}_{\mathrm{full}}$.

Before we compute this ratio, let's prove a combinatorial relation that will help us in the next steps. This relation is
\begin{equation}\label{comb_relation}
    \sum^{K}_{k=0} \binom{k+r}{r} = \binom{K+r+1}{r+1}.
\end{equation}

Following the induction principle, we straightforwardly have that, for the base case of $K=0$, the previous equation is true. Thus, we now assume the result for $K$, and verify it is valid for $K+1$. Hence,
\begin{equation}
    \sum^{K+1}_{k=0} \binom{k+r}{r} = \binom{K+1+r}{r} + \sum^{K}_{k=0} \binom{k+r}{r} = \binom{K+1+r}{r} + \binom{K+1+r}{r+1}.
\end{equation}
\noindent According to the Pascal's Triangle~\cite{cameron1994combinatorics}, that is
\begin{equation}
    \binom{A}{B} + \binom{A}{B-1} = \binom{A+1}{B},
\end{equation}
\noindent making $A=K+r+1$ and $B=r+1$, we obtain
\begin{equation}
    \sum^{K+1}_{k=0} \binom{k+r}{r} = \binom{K+1+r}{r} + \binom{K+1+r}{r+1} = \binom{K+r+2}{r+1} = \binom{(K+1) + r +1}{r+1},
\end{equation}
\noindent where the last step is just to make clear that Eq.~\eqref{comb_relation} holds, concluding our demonstration.

With this result in hand, we now analyze the term $N^{(\mathrm{i})}_{\mathrm{full}}$, which yields
\begin{equation}
    N^{(\mathrm{i})}_{\mathrm{full}} = \sum^{N}_{n_0=0} N Q^{(n_0)}_S(N,M) = N \sum^{N}_{n_0=0} \binom{N-n_0+M-2}{M-2} = N \sum^{N}_{k=0} \binom{k+M-2}{M-2} = N\binom{N+M-1}{M-1},
\end{equation}
\noindent where we have made the changes of variables $k=N-n_0$ and $r=M-2$ to use the result proved above. The result of the previous equation is expected because the sums for $N^{(\mathrm{i})}_{\mathrm{full}}$ are equivalent to the combinatorial problem of distributing $N$ indistinguishable photons in $M$ modes, which is given by the binomial coefficient $\binom{N+M-1}{M-1}$. Moving forward, similarly for $N^{(\mathrm{f})}_{\mathrm{full}}$, we have
\begin{equation}
    N^{(\mathrm{f})}_{\mathrm{full}} = \sum^{N}_{n_0=0} \left(N-n_0\right) \binom{N-n_0+M-2}{M-2} = \sum^{N}_{k=0} k \binom{k+M-2}{M-2}.
\end{equation}
\noindent However,
\begin{equation}
    k \binom{k+M-2}{M-2} = \frac{k (k+M-2)!}{(M-2)!k!} = \frac{(M-1) (k+M-2)!}{(M-1)!(k-1)!} = (M-1)\binom{(k-1) + (M-1)}{M-1},
\end{equation}
\noindent leading to
\begin{equation}
     N^{(\mathrm{f})}_{\mathrm{full}}  = \sum^{N}_{k=0}(M-1)\binom{(k-1)+ (M-1)}{M-1} = (M-1)\binom{N+M-1}{M},
\end{equation}
\noindent Finally, computing the ratio we have,
\begin{equation}\label{photon ratio}
    \frac{N^{(\mathrm{f})}_{\mathrm{full}}}{N^{(\mathrm{i})}_{\mathrm{full}}} = \frac{(M-1)\binom{N+M-1}{M}}{N\binom{N+M-1}{M-1}} = \frac{ \frac{(M-1)(N+M-1)!}{(N-1)!M!} } { \frac{N(N+M-1)!}{N!(M-1)!} } = \frac{(M-1)(N+M-1)!N!(M-1)!}{(N-1)!M!N(N+M-1)!}  = \frac{M-1}{M}.
\end{equation}

\begin{figure*}[h!]
\includegraphics[width=1.0\linewidth]{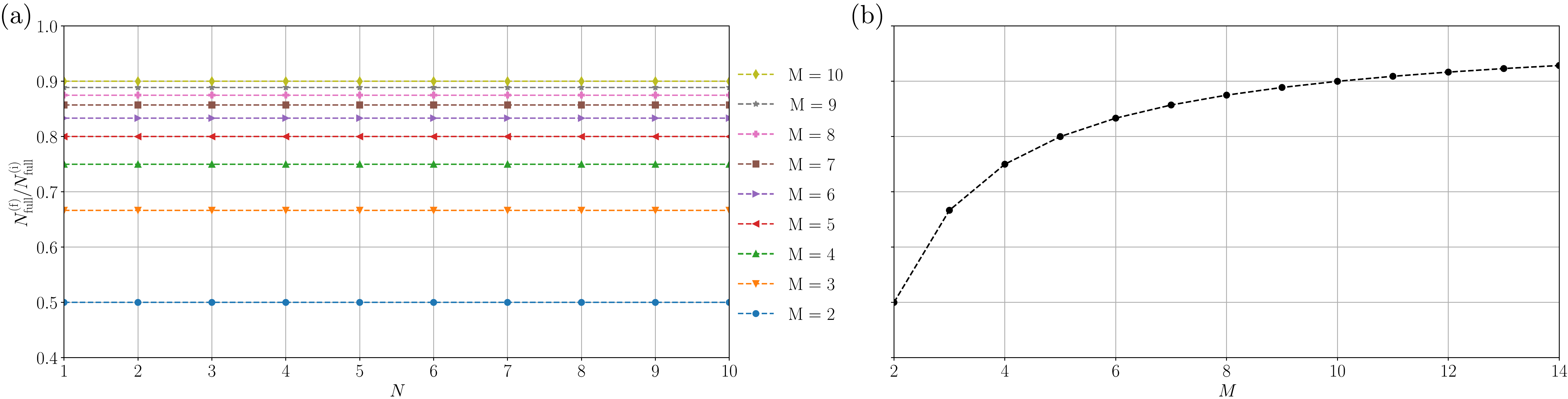}
	\caption{(a) The ratio of the final to the initial total number of excitations as a function of total photons $N$ for different numbers of modes -- see the legend. (b) The ratio of the final to the initial total number of excitations as a function of number of modes $M$, assuming a fixed number of excitations $N=5$. In both plots, we consider the general result presented in Eq.~\eqref{ratio sums}. Besides the independence with respect to $N$, we observe that as the number of modes increases, the remaining energy in the antisymmetric modes also scales up, and verify the analytical result of Eq.~\eqref{photon ratio}.}
	\label{ratio}
\end{figure*}

Hence, the ratio of the final to the initial number of excitations for a set of collective states considering a fixed total number of excitations $N$ in the $M$ modes depends only on the number of modes interacting with the atom, independent of the number of photons. Moreover, as the number of modes increases, the number of photons in the modes after the dynamics reaches the steady state also increases. Following our previous discussions, the photons of the steady state are present in the antisymmetric modes, being inaccessible to us as long as the symmetry of the problem does not break. Thus, there always exists a number of photons that we do not detect since they are in such states that do not interact with matter. In Fig.~\ref{ratio}, we plot for some values of $M$ and $N$ the ratio values considering the sums presented for the general result of Eq.~\eqref{ratio sums}, which confirms our demonstration.

This last result can be understood as the combinatorial analysis problem of distributing $N$ objects (photons) into $M$ shelves (modes). Since photons and modes are indistinguishable, the different ways to distribute such photons are irrelevant and then can be treated as a comminatory of identical objects in any order. In this way, the number of states with $n_\mu$ photons in the $\mu$-th collective mode is the same and independent of the mode we are considering. Therefore, the number of photons in each collective mode is the same ($N/M$) so that, when we sum over the $M$ modes, we have the total number of excitations $M (N/M) = N$. However, since only the photons of the symmetric mode are dissipated, only a fraction of photons are exchanged by the thermal sources. This fraction corresponds to the number of photons in the symmetric mode, $N/M$. In other words, $N(M-1)/M$ photons will remain in the bosonic modes, precisely in the antisymmetric modes. Thus, when we divide the final number of total excitations ($N^{(\mathrm{f})}_{\mathrm{full}} = N(M-1)/M$) by the initial number of total excitations ($N^{(\mathrm{i})}_{\mathrm{full}} = N$), we achieve the ratio of $(M-1)/M$. This result reveals the symmetry of the problem since the number of photons does not change how they are distributed in the $M$ collective modes, making the analysis done here for the first collective mode equally applied to the other modes.

\section*{The detectable and undetectable intensity}

The intensity can be calculated as~\cite{Glauber1963_1}
\begin{equation}
    I = \langle E^-E^+\rangle.
\end{equation}
\noindent Thus, in the collective basis, it becomes $I = M\langle \hat{A}^\dagger_0 \hat{A}_0\rangle$, where it is evident the detection only of the photons of the symmetric mode, as expected, but with enhancement proportional to number of modes, which conceals the reduced detection. As discussed above, after the calculation of the photon ratio, just a fraction of $1/M$ of all excitations is detected. However, it is detected $M$ times stronger due to the collective effect, compensating the energy not detected from the dark photons, i.e., the photons in dark states. Comically, while collective effects result in the measurement of photons in only a specific mode, these same collective effects simultaneously cause this fraction of photons to couple more strongly with matter.

Considering $M=2$, the electrical field operator present in the Jaynes-Cummings interaction Hamiltonian is $E^+ = a_1+a_2$ -- with no relative phase for simplicity -- which is proportional to the symmetric mode. However, we can change the symmetry of the problem in such a way that the bright states become the dark states and vice versa. To verify it is easy since, in two modes, the symmetric operator is $\hat{A}_0 = a_1+a_2$, and the antisymmetric operator is $\hat{A}_1 =a_1-a_2$. Thus, if the Fock operators are in phase, their collective operator is the symmetric one, and if they have opposite phases, their collective operator is the antisymmetric one. Therefore, if one is able to construct such an interaction ruled by a hypothetical electric operator given by $\tilde{E}^+ = a_1-a_2$, this is the same as $E^+=\sqrt{M}\hat{A}_1$. Hence, the intensity related to the antisymmetric operator is $I = \langle \tilde{E}^-\tilde{E}^+\rangle = M \langle \hat{A}_1^\dagger \hat{A}_1 \rangle$, which is proportional only to the number of photons of the antisymmetric mode. In this sense, the intensity related to the photons present in the antisymmetric modes that was undetectable now becomes detectable given the new interaction. In addition, by symmetry, one can expect that the photons are equally distributed in the $M$ collective modes. Eq.~\eqref{photon ratio} reveals this symmetry once the result is independent of $N$ and is equal to the ratio of the number of antisymmetric to the total number of modes. Thus, for $M>2$, there are only $1/M$ photons in the symmetric mode, which increases the undetectable intensity.

As an example, considering the result of Eq.~\eqref{photon ratio}, which basically tells us the fraction of photons that are initially in antisymmetric modes, if we are in a room with 100 lamps (considering that each one emits light at the same frequency), we would detect only $1\%$ of the photons emitted, while $99\%$ of the photons would be undetectable. Intriguing, if we try to measure the intensity produced by the same set of lamps, it will result in $100\%$ of their power. This occurs because the apparatus used to measure the intensity interacts only with the symmetric mode, which has an enhancement in the coupling strength of $\sqrt{M}$, as one can see from the electrical operator when written in the collective basis. Since the intensity is proportional to the square of the electric field, we measure each one of the $1\%$ of the photons emitted by the lamps $100$ times stronger, which makes up for the reduced number of detectable photons.

\section*{Non-resonant case}

\begin{figure}[h!]
\includegraphics[width=0.5\linewidth]{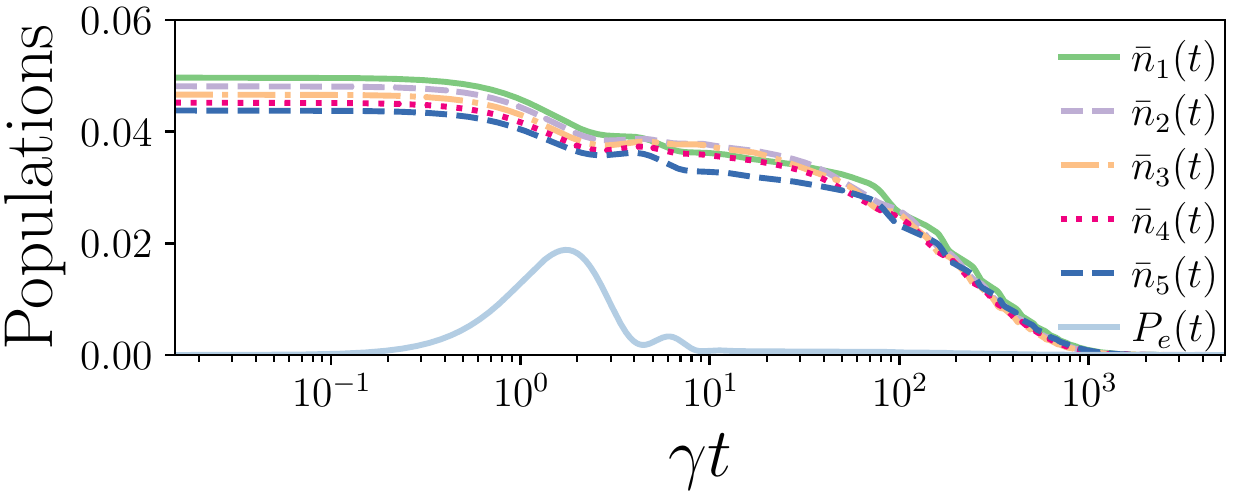}
	\caption{Number of photons in the bare modes (see legend) and the population of the excited state of a dissipative two-level atom ($P_e(t)$) as functions of the dimensionless time $\gamma t$. In this plot, $\gamma/g = 3$ and the initial average photon number for the first bare mode is $\bar{n}_1 = 0.05$. The frequency detuning between the $j+1$-th and the $j$-th bare mode is $\delta=\gamma/15$, with $1\leq j \leq 4 \in \mathbb{N}$. The bare modes are initially at thermal states, at the same temperature and different average photon numbers (since they have different frequencies). Still, the atom is in its ground state initially, and we assume the atomic frequency resonant with the third atom. In this case, all the initial energy stored in the bare modes is exchanged and dissipated.}
	\label{nonresonant_case}
\end{figure}

When considering $M$ modes, now assuming that they have different frequencies, the electrical field and the Hamiltonian will depend on these frequencies in such a way that the collective states become functions of phase relationships between the modes~\cite{diniz2024}. In this context, the states that either strongly (bright) or do not (dark) couple to the matter vary in time. At a specific instant of time, given a number of excitations $N$, there is one bright state and a fixed number of intermediate states that exchange photons with the dissipative atom, similarly to the resonant case. In the same way, the number of dark states is also fixed. However, at the subsequent instant of time, the state that was bright will no longer have this characteristic, and equivalently for the other states. As a consequence, the dissipative atom is able to interact with all the collective states but at different times. Therefore, all the photons (energy) will be exchanged and dissipated. As a possible application, making such time dependencies appear can be used by one to break the symmetry when dealing with the resonant case in order to be able to access the energy of the dark states~\cite{Diniz_2025_reset}.In Fig.~\ref{nonresonant_case}, we plot the interaction of five non-resonant bare modes with a dissipative two-level atom, where we note the dissipation of the entire energy initially stored in the modes.

\end{document}